\documentclass[aps,pre,twocolumn,superscriptaddress,natlib,showpacs]{revtex4-1}
\usepackage{natbib}
\usepackage{graphicx}
\usepackage{dcolumn}
\usepackage{bm}
\usepackage{color}
\usepackage{amssymb}
\usepackage{amsmath}
\usepackage{amscd} 
\begin{document}

\title{
Hydrodynamic memory can boost enormously driven nonlinear diffusion and transport }

\author{Igor Goychuk}
\email{igor.goychuk@fau.de, corresponding author}
\affiliation{Institute for Multiscale Simulation, Friedrich-Alexander University 
of Erlangen-Nueremberg, Cauerstr. 3,  91058 Erlangen, Germany}
\author{Thorsten P\"oschel}
\affiliation{Institute for Multiscale Simulation, Friedrich-Alexander University 
of Erlangen-Nueremberg, Cauerstr. 3,  91058 Erlangen, Germany}

\date{\today}

\begin{abstract}

Hydrodynamic memory force or Basset force is known since the 19th-century. Its influence on
Brownian motion remains, however, mostly unexplored. Here, we investigate its role in
nonlinear transport and diffusion within a paradigmatic model of tilted washboard potential.
In this model, a giant enhancement of driven diffusion over its potential-free limit 
[Phys. Rev. Lett. \textbf{87}, 010602 (2001)] presents
a well-established paradoxical phenomenon. In the overdamped limit, it occurs at a critical
tilt of vanishing potential barriers. However, for weak damping, it takes place surprisingly
at another critical tilt, where the potential barriers are clearly expressed. Recently we
showed [Phys. Rev. Lett. \textbf{123}, 180603 (2019)] that Basset force could make 
such a diffusion enhancement enormously large. In this
paper, we discover that even for moderately strong damping, where the overdamped theory works
very well when the memory effects are negligible, substantial hydrodynamic memory
unexpectedly makes a strong impact.  First, the diffusion boost occurs at non-vanishing
potential barriers and can be orders of magnitude larger. Second, transient anomalous
diffusion regimes emerge over many time decades and potential periods. Third, particles'
mobility can also be dramatically enhanced, and a long transient super-transport regime
emerges.

\end{abstract}

\maketitle

\section{Introduction}

Non-Markovian hydrodynamic memory effects due to Boussinesq-Basset force
emerging in the motion of macroscopic bodies with fluctuating velocity are known
since the 19th-century \cite{Boussinesq, Basset, LandauHydro} and still present
an active field of research \cite{Druzhinin94,Guseva13}. Their influence on
Brownian motion \cite{Frey05, RiskenBook, CoffeyBook,HanggiRevModPhys} remains,
however, largely unexplored, even though they are known to effect famous
algebraic tail in the velocity autocorrelation function (VACF)  of Brownian
particles \cite{Widom71, KuboBook}, which was found first in molecular dynamic
simulations by Alder and Wainwright \cite{AlderPRL}. Neither diffusion, nor
transport are, however, affected asymptotically in the absence of nonlinear
force fields, even if a transient superdiffusion is engendered
\cite{MainardiFLE}. The existence and importance of such memory effects  was
experimentally manifested for free diffusion of colloidal particles
\cite{Weitz89}, and,  more recently, for particles trapped in parabolic
potentials \cite{FranoschNature,HuangNatPhys,Kheifets14}. It raises the question
of their general role and importance in nonlinear transport and diffusion
\cite{GoychukPRL19}, where the model of tilted washboard potentials serves as a
paradigm in condensed matter physics and far beyond \cite{RiskenBook,
CoffeyBook,HanggiRevModPhys}. Within this model, a giant enhancement of driven
diffusion \cite{Costantini1999, ReimannPRL01, LindnerFNL01} over its
potential-free limit is a well-established paradoxical phenomenon in the
overdamped limit \cite{LeeGrierPRL, EvstigneevPRE, CoffeyBook}, where the
inertial effects are entirely negligible. It occurs at a critical potential tilt
of vanishing potential barriers \cite{ReimannPRL01, LindnerFNL01, CoffeyBook}
with applied constant force $f_c^{(1)}=1$ in the units used in this paper.
Inertial effects in nonlinear diffusion beyond thermal equilibrium are less
studied \cite {RiskenBook, CoffeyBook} and brought many surprises
\cite{SanchoPRL04, LindenbergPRL, Marchenko12, Marchenko14, LindnerSokolovPRE}
lately. However, the influence of hydrodynamic memory effects on such a
nonlinear driven diffusion and transport came only recently in the spotlight of
attention \cite{GoychukPRL19}.  For weak damping, a giant enhancement of
diffusion occurs at another critical tilt, where the potential barriers are
clearly expressed \cite{Marchenko12, Marchenko14, LindnerSokolovPRE}. Then,
profound memory effects make such an enhancement enormously large and result in
a substantial prolongation of a transient superdiffusion regime
\cite{GoychukPRL19}. 

To understand the mechanism of a resonancelike enhancement of diffusion in the
corresponding weakly-damped regime, the picture of motion bistability originally
developed by Risken and collaborators \cite{RiskenBook,Vollmer1980} in the
absence of hydrodynamic memory becomes crucially important. There exists a
critical friction value $\gamma_0^{(c)}\approx 1.193$~\cite{RiskenBook} (in
scaled units used below) such that for a smaller friction at zero temperature
there emerges a friction-dependent critical tilt value $f_c^{(3)}$ (our
notations are different from used in \cite{RiskenBook}) such that for tilting
forces between two critical values, $f_c^{(3)}<f<f_c^{(1)}$ the motion is
bistable at zero temperature with thermal fluctuations neglected. 
An excellent account of this bistability is given in \cite{Marchesoni97}
in the context of stochastic resonance problem.
Depending on
the starting point in the phase space, any particle will either end  in one of
potential wells due to frictional losses, or run indefinitely, when losses are
compensated by the energy delivered from the external field. In this bistable
regime and at a finite temperature $T$, velocity of particle exhibits bistable
fluctuations for a sufficiently small friction $\gamma_0\ll \gamma_0^{(c)}$.
They can be modeled and well understood as thermally activated fluctuations in a
bistable velocity pseudo-potential $V(v)=-k_BT\ln[P(v)]$, where $P(v)$ is
velocity distribution and $k_B$ is Boltzmann constant
\cite{Marchenko14,Marchenko2017,LindnerSokolovPRE}.  One potential minimum
corresponds to $v_1=0$ (trapped particles), and another one to $v_2=f/\gamma_0$
(running particles, units are scaled). Near minima, $P(v)$ is approximately
Gaussian (Maxwell distribution) with a thermal velocity width
\cite{Marchenko14,Marchenko2017}. The giant enhancement of diffusion occurs at
the condition of equal pseudo-potential minima or equal probabilities
\cite{Marchenko14,LindnerSokolovPRE} for the particle to be trapped or run in an
intermittent L\'evy walk like fashion
\cite{Geisel87,Zumofen93,Shlesinger93,SanchoPRL04}.  Such a critical condition
can also be obtained in a generic model of velocity-bistable active diffusion
\cite{LindnerPRL08}. For underdamped dynamics in washboard potential, the
diffusion maximum occurs at some $f_c^{(2)}$, $f_c^{(3)}< f_c^{(2)}<f_c^{(1)}$
satisfying this condition. 

Upon taking hydrodynamic memory influence for a small $\gamma_0$ into account,
this basic picture remains approximately valid upon some essential modifications
\cite{GoychukPRL19}. First, the diffusion enhancement becomes strongly amplified
and sharpened (suppressed outside of the narrow maximum region). Second, the
distribution of particle velocities in the running state (near its maximum) is
broader than Maxwellian. This effect can be characterized by either enhanced
kinetic temperature in the running state, or by a smaller effective mass of the
particle in this state. The latter interpretation is preferred because it is
convenient to characterize the whole velocity distribution by a kinetic
temperature measured by its width
\cite{Brilliantov04,SieglePRL10,SiegleEPL11,Marchenko12,GoychukPRL19}. Particles
become kinetically hot in the bistable regime. Third, an effective friction
experienced by particles becomes enlarged by the hydrodynamic memory friction.
This leads to an effective suppression of the asymptotically normal transport in
comparison with the memoryless case. However, transient regime of looking
anomalously fast transport, $\langle \delta x(t)\rangle\sim t^{\kappa_t}$, with
$\kappa_t>1$ can be prolonged enormously. Likewise, transient superdiffusion,
$\langle \delta x^2(t)\rangle\sim t^{\kappa_d}$, with $\kappa_d>1$ can also be
drastically prolonged in time. It is due to the changed kinetics of the
transitions between two macrostates of velocity  L\'evy walk, which becomes
anomalously slow, stretched-exponential, instead of exponential in the
memoryless case -- the fourth profound feature introduced by hydrodynamic
memory. Furthermore, like in the case of asymptotically superdiffusive transport
\cite{SieglePRE10,SieglePRL10,SiegleEPL11}, long hyperdiffusive regimes,
$\kappa_d>2$, are present due to transiently growing in time kinetic
temperature. In the memoryless case, such regimes are also present
\cite{Marchenko12}. However, they are much shorter.         

Now, novel profound questions emerge: How these interesting features introduced
by hydrodynamic memory are modified beyond the $\gamma_0\ll  \gamma_0^{(c)} $
regime  studied in \cite{GoychukPRL19}? Is hydrodynamic memory still important
for $\gamma_0$ equal and even larger than Risken's $\gamma_0^{(c)}$, when
dynamics becomes overdamped? This question is very important because Brownian
motion in fluids is typically overdamped. For example, in experimental works
\cite{FranoschNature,Kheifets14} colloidal particles are overdamped and,
nevertheless, exhibit resonances caused by the hydrodynamic memory. Next, does
hydrodynamic memory always increase an effective friction or it can also make
that smaller, e.g. for a sufficiently large $\gamma_0$, and how large is large?
For example, some results in recent Ref. \cite{Seyler19} for transport in a
critically tilted piecewise linear  periodic potential at $T=0$ imply that this
can be the case.  Next, remains picture of bistable velocity fluctuations valid
for sufficiently large $\gamma_0$, which is yet smaller than $\gamma_0^{(c)}$?
Actually, some results presented in the Supplemental Material \cite{supplPRL19}
of Ref. \cite{GoychukPRL19} imply ``no'' already for $\gamma_0$ larger than
about $0.25$, which was confirmed in a recent detailed study
\cite{Spiechowicz20}. This feature means that the running velocity state is not
necessarily monostable unless $\gamma_0$ is small enough.  The numerical
simulations reveal that already for $\gamma_0=0.3$, the velocity distribution
can be trimodal, see Fig. 7, (a) in \cite{supplPRL19}, and, especially, the
panel (c) therein, for $\gamma_0=0.7$, where the running state consists, in
fact, of two velocity substates with $P(v)$ maxima at $v_{2}^{(1)}$ and
$v_{2}^{(2)}$ such that $v_{2}^{(1)}<v_2=f/\gamma_0<v_{2}^{(2)}$. Moreover,
$v_2$ corresponds to the minimum (!) and not maximum of $P(v)$, as bistable
picture of $V(v)$ \cite{Marchenko12,Marchenko14}, valid for sufficiently small
$\gamma_0$ only \cite{GoychukPRL19,Spiechowicz20}, can misleadingly imply. For a
critical tilt $f=1$ in Fig. 7, (d) in \cite{supplPRL19}, the minimum at $v=0$
(trapped state) disappears, and the \textit{running} state remains bistable. It
means that velocity fluctuations can remain bistable even for an overcritical
tilt, when the trapped states are absent. Hence, the case of nearly overdamped
dynamics is not trivial, even if to neglect crucial memory effects.

Below we show that even for a moderately strong damping, within a seemingly
overdamped regime, hydrodynamic memory unexpectedly makes a very profound
impact. As a general implication, it means that hydrodynamic memory effects,
whose neglect might earlier seem intuitively be well justified, can nevertheless
profoundly affect nonlinear transport and diffusion. They should not be
generally \textit{ad hoc} neglected in further research. Theory of nonlinear
Brownian motion in fluids should be rethought and revisited from this angle of
view.

\section{Model and Theory} We consider one-dimensional transport and diffusion of spherical
Brownian particles with radius $R$ and mass $m=4\pi \rho R^3/3$ ($\rho$ is the particles' mass
density) in a fluid with kinematic viscosity $\mu$ and density $\rho_f$ governed by a fractional
Langevin equation (FLE)
\cite{MainardiFLE,Lutz01,CoffeyBook,SiegleEPL11,GoychukACP12,FranoschNature}  
\begin{eqnarray} \label{FLE1}
m^*\ddot x(t) +\eta_0 \dot x(t)+  \eta_{\alpha}\sideset{_{-\infty}}{_t}{\mathop{\hat D}^{1/2}}
\dot x(t)\\ \nonumber 
=f(x)+ \xi_0(t)+\xi_{\alpha}(t)\;    
\end{eqnarray}  in a periodic
force-field 
\begin{equation} 
f(x)=f_c^{(1)}\sin(x/x_0)+f
\label{field} 
\end{equation} 
with amplitude
$f_c^{(1)}=U_0/x_0$. Here, $U_0$ is the amplitude of the corresponding washboard potential with
period $L=2\pi x_0$, which is biased by a constant driving force $f$. Trapped states exist only
below the critical value $f_c^{(1)}$, $f<f_c^{(1)}$. In Eq. (\ref{FLE1}), $m^*=m+2\pi \rho_f
R^3/3$ is a fluid-renormalized mass of a Brownian particle \cite{LandauHydro,KuboBook,ChaikinBook},  
$\eta_0=6\pi R\rho_f\mu$ is Stokes viscous friction, $\sideset{_{-\infty}}{_t}{\mathop{\hat
D}^{\alpha-1}}v(t):=\frac{1}{\Gamma(2-\alpha)} \frac{d}{dt}\int_{-\infty}^t dt'
v(t')/(t-t')^{\alpha-1}$, with $\alpha=3/2$, is Riemann-Liouville fractional derivative
\cite{MainardiFLE,Mathai17}, and $\eta_\alpha=\eta_0\sqrt{\tau_r}$  is a  fractional friction
coefficient. The corresponding memory term in the FLE reflects hydrodynamic memory, which is
characterized by a relaxation time scale $\tau_r=R^2/\mu$ entering $\eta_\alpha$.  It presents
the Boussinesq-Basset force, which is derived within similar approximations as the Stokes
friction, however, for a particle with fluctuating  velocity (non-steady Stokes
flow) \cite{Boussinesq,Basset,LandauHydro}. 
Roughly speaking, $\tau_r$ is a characteristic time for a backflow induced by the body motion to diffuse over its size. Inertial effects in the particle's dynamics are also not always negligible. 
The characteristic velocity relaxation time, $\tau_v=m^*/\eta_0$ (obtained in neglecting the memory effects), 
is $\tau_v=\tau_r (2\rho/\rho_f+1)/9$, 
in terms of $\tau_r$ and the ratio $\rho/\rho_f$ of the body and fluid densities. For example, 
in the case of a neutrally buoyant particle, $\rho_f=\rho$, $\tau_v=\tau_r/3$, and $\tau_v=\tau_r$ at $\rho=4\rho_f$. It means that unless the Brownian particle is very heavy with respect to fluid, hydrodynamic memory is not neglible once the particle's inertia becomes important, especially given a slow algebraic character of this memory decay. We recast the Boussinesq-Basset force in the form of memory friction \cite{MainardiFLE, SiegleEPL11,
GoychukACP12, GoychukPRL19} , $\int_{-\infty}^{t}\eta(t-t')\dot
x(t')dt'$, with a singular memory kernel $\eta(t)$ corresponding to the operator of the
Riemann-Liouville fractional derivative. For $1<\alpha<2$, $\eta(t>0)=-\eta_\alpha
t^{-\alpha}/|\Gamma(1-\alpha)|<0$, however, $\int_0^t  \eta(t')dt'\sim t^{1-\alpha}>0$ is always
positive and tends to zero with  $t\to\infty$. This term is absent for $v=\dot x=const$.
However, it is always  present in the realm of Brownian particles, where it must be complemented
by  the corresponding unbiased thermal Gaussian force $\xi_\alpha(t)$ obeying the 
fluctuation-dissipation relation (FDR) \cite{Kubo66,KuboBook}, $\langle 
\xi_\alpha(t')\xi_\alpha(t)\rangle=k_BT\eta(|t-t'|)$, which follows from the  fundamental
fluctuation-dissipation theorem (FDT) \cite{Kubo66,KuboBook}. 
$\xi_\alpha(t)$  provides a naturally emerging instance of the
fractional Gaussian noise or fGn \cite{Mandelbrot68}. By the same token, $\langle 
\xi_0(t')\xi_0(t)\rangle=2k_BT\eta_0\delta(|t-t'|)$, as in the standard  Langevin equation,
where $\xi_0(t)$ is a white Gaussian noise, which like fGn is a singular stochastic process with
infinite variance existing only in a  class of distributions. FLE (\ref{FLE1}) presents an
important example of general nonlinear Generalized Langevin Equation or GLE \cite{Kubo66,
KuboBook, GoychukACP12}.

Periodic potentials acting on  Brownian microparticles can be created by a lattice of optical
vortices \cite{LeeGrierPRL} or optical tweezers \cite{EvstigneevPRE} (with $L$ in the sub-micron
range), or, e.g.,  by nanoimprint lithography \cite{Guo}, for nanoparticles (down to nanometer
scale).  The FLE description was confirmed experimentally for colloidal particles in parabolic
traps \cite{FranoschNature, HuangNatPhys, Kheifets14}, where hydrodynamic effects were measurable and even
caused resonances in the case of almost overdamped dynamics \cite{FranoschNature}. They,
however, never were studied for nonlinear Brownian transport and diffusion until recently
\cite{GoychukPRL19}, even theoretically,  except for a model case, where the Stokes friction was
\textit{ad hoc}  neglected \cite{SiegleEPL11}. Indeed, in the case of potential-free diffusion,
$U_0=0$, the memory effects do not affect the diffusion coefficient,  $D_0=k_BT/\eta_0$,
asymptotically. However, they do cause some relatively short transient
superdiffusion \cite{MainardiFLE} and profoundly modify the stationary VACF, $\langle
v(t)v(0)\rangle_{\rm st}$. Namely, it universally acquires asymptotically a long algebraic
tail,  $\langle v(t)v(0)\rangle_{\rm st}\sim v_T^2 
\sqrt{\tau_r/\pi}/(2\gamma_0t^{3/2})$ \cite{Widom71, MainardiFLE, GoychukPRL19},  where
$v_T=\sqrt{k_BT/m^*}$ is thermal velocity and $\gamma_0=\eta_0/m^*$. This tail has first been
found in molecular-dynamics simulations by  Alder and Wainwright\cite{AlderPRL}. Moreover, the
initial decay of VACF is stretched-exponential and not exponential, in the case of strong yet
realistic memory effects \cite{GoychukPRL19}. The case of driven nonlinear diffusion is capable
of further surprises \cite{GoychukPRL19}.  

We shall scale distance  in $x_0$, time in $\tau_0=x_0\sqrt{m^*/U_0}$, which is inverse circular
 frequency of oscillations at the bottom of potential wells in the absence of friction and bias,
 energy in $U_0$, and temperature as $\tilde T=k_BT/U_0$. In these units, $f_c^{(1)}=1$ and
 dimensionless $\tilde \gamma_0=\gamma_0\tau_0$ (tilde will be mostly omitted in the following)
 measures the strength of normal friction. For $\eta_\alpha=0$, the unbiased intrawell dynamics
 is overdamped for $\gamma_0\geq 2$.  Furthermore, $\gamma_\alpha=\eta_\alpha/m^*$ in these
 units reads $\gamma_\alpha=3\sqrt{\gamma_0/(1+2\rho/\rho_f)}$, which is maximal,
 $\gamma_\alpha^{\rm (max)}=3\sqrt{\gamma_0}$, in the limit of ultralight particles,
 $\rho/\rho_f\to 0$. The memory effects are fully negligible in the opposite limit
 $\rho/\rho_f\to \infty$, and are expected to be strong for $\rho \sim \rho_f$ or smaller. FLE
 does not allow for analytical solutions for the considered nonlinear dynamics and we solved it
 numerically \cite{GoychukPRL19}, as detailed in the Appendix A.

\section{Results and Discussion}

\begin{figure}
\centering
	\includegraphics[width=.8\linewidth]{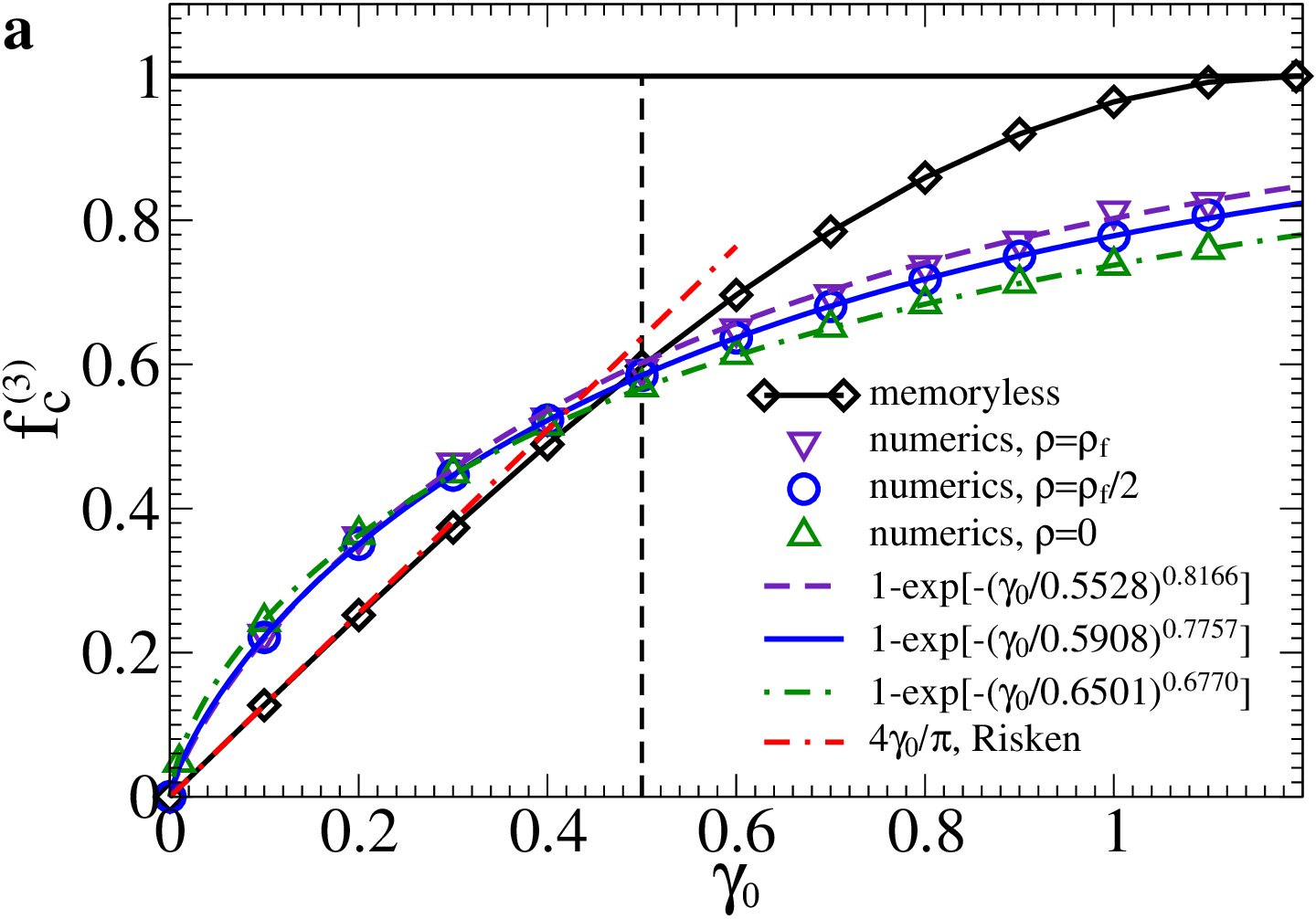}\\
	\vspace{0.8cm}
	\includegraphics[width=.8\linewidth]{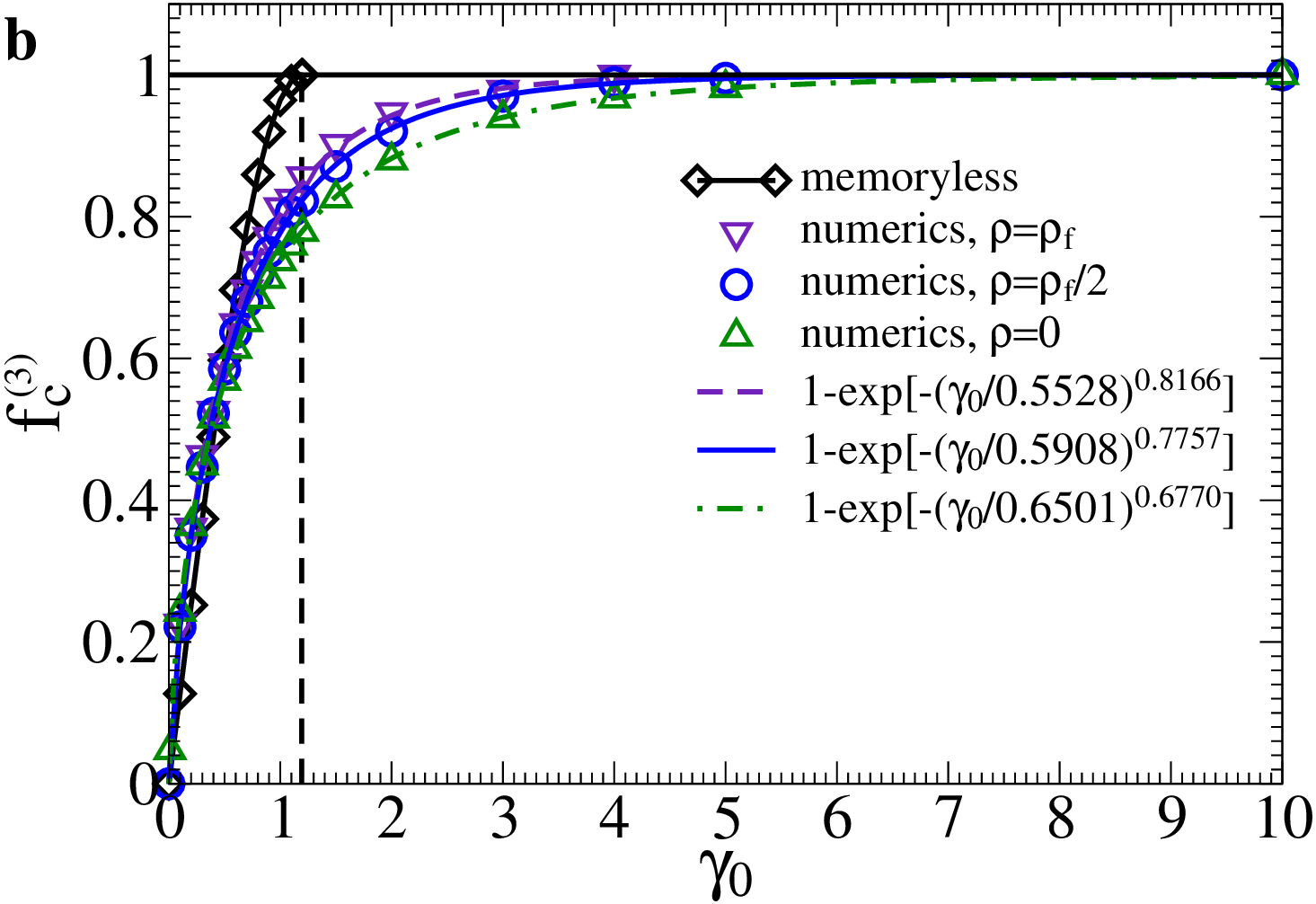}
	\caption{Phase diagram of bistability at $T=0$. Dependence of the
		critical force $f_c^{(3)}$ on $\gamma_0$ without and in the
		presence of hydrodynamic memory effects for three fixed values
		of ratio $\rho/\rho_f$ shown in the plot. For $f<f_c^{(3)}$,
		there are no running states. Every trajectory eventually ends in
		a potential well (trapped solutions). For $f_c^{(3)}<f<
		f_c^{(1)}=1$, running trajectories co-exist with trapping
		solutions, and for $f>1$ the only running solutions remain. In
		the memoryless case \cite{RiskenBook},  $f_c^{(3)}\approx
		4\gamma_0/\pi$, for $\gamma_0 \lesssim 0.25$, see
		red double-dash-dotted line in panel (a).  For
		for $\gamma_0 \lesssim 0.5$, memory effects increase the
		effective friction, which can be  judged upon the
		correspondingly increased $f_c^{(3)}$. This trend is changed to
		the opposite for  $\gamma_0>0.5$. The vertical line in panel (a)
		at $\gamma_0=0.5$ helps to realize this. Moreover, the regime of
		bistability extends far beyond  the Risken's
		$\gamma_0^{(c)}\approx 1.193$  [see  vertical line in panel
		(b)], as panel (b) manifests.}
\label{Fig1}
\end{figure}

\subsection{Influence of memory effects on bistability phase diagram}

\begin{figure}
\centering
  \includegraphics[width=.8\linewidth]{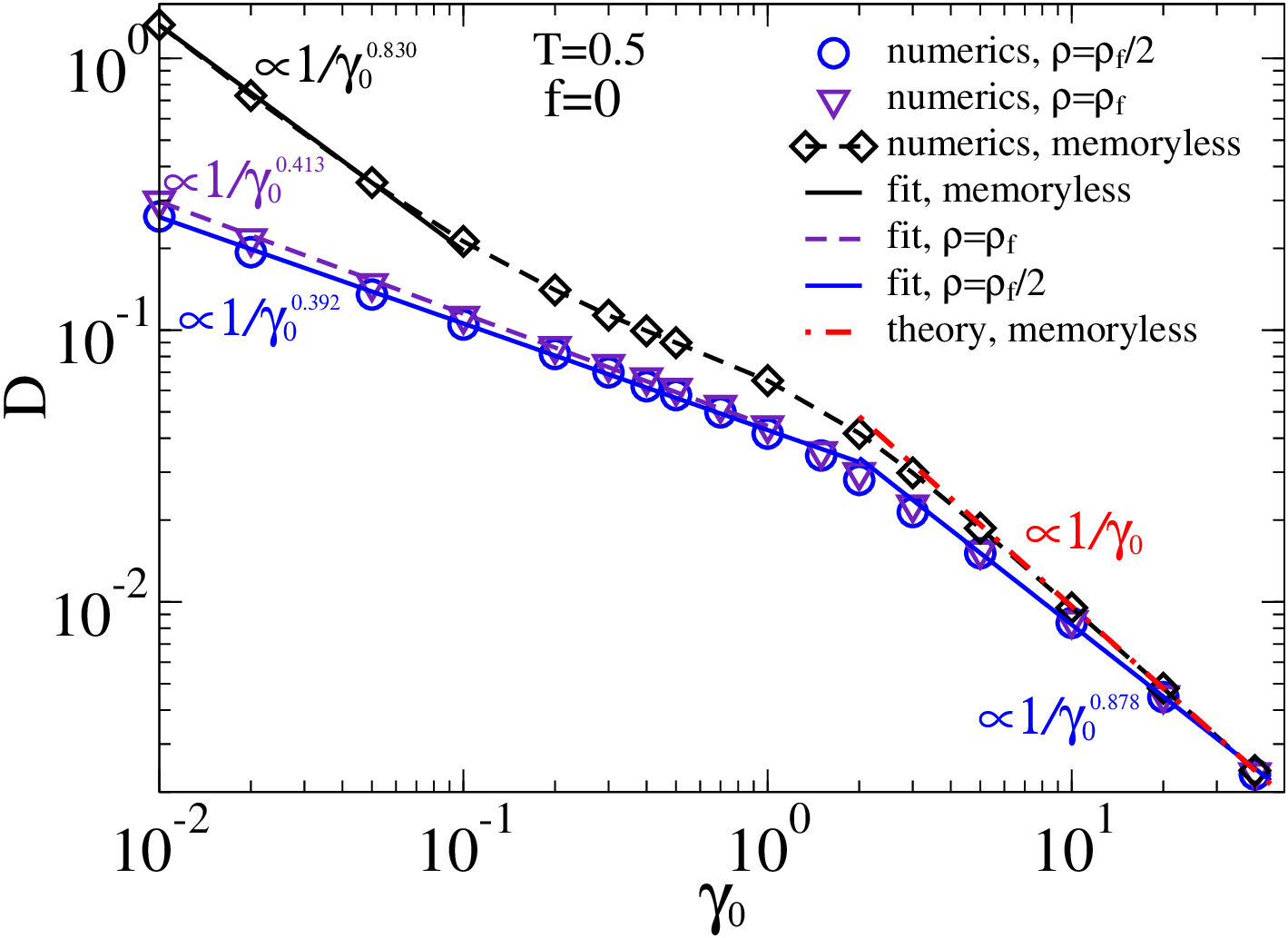}
\caption{Dependence of nonlinear diffusion coefficient on 
viscous friction in the case of unbiased diffusion at $T=0.5$.
 Two values of the ratio of particles 
to fluid density are considered, $\rho/\rho_f=1$ and $\rho/\rho_f=1/2$, as well as the memoryless 
case, for comparison, see major text for detail. Numerical data are shown 
by indigo triangles for $\rho=\rho_f$ 
(with  a small-$\gamma_0$ fit depicted by dashed indigo line coming through the symbols), blue circles for 
$\rho=\rho_f/2$ (with small and large $\gamma_0$ fits depicted by full blue lines coming through 
the symbols), and dashed 
black line with diamonds, in the case of memoryless dynamics. Analytical result for a 
large-friction limit of memoryless dynamics is depicted by dash-dotted red line, and a small-$\gamma_0$ fit
to this dynamics by full black line.}
\label{Fig2}      
\end{figure}

As mentioned in Introduction, the fundamental feature of driven underdamped dynamics
in tilted washboard potential in the absence of memory effects is its bistability
\cite{RiskenBook}. Namely, for $\gamma_0$ less than a critical value
$\gamma_0^{(c)}\approx 1.193$ \cite{RiskenBook}, there exists a critical force
$f_c^{(3)}<f^{(1)}_c$ such that for $f<f_c^{(3)}$, any trajectory in the phase space
will end eventually at $T=0$ in one of the potential wells. However, for
$f_c^{(3)}\leq f< f^{(1)}_c$, the deterministic running solutions emerge and co-exist
with the trapped ones, whereas for $f>f^{(1)}_c$, only the running solutions remain.
This critical force value depends on $\gamma_0$. For a small friction,
$f^{(3)}_c\approx 4\gamma_0/\pi$,  see Fig. 11.26 in Ref. \cite{RiskenBook} and
red double-dash-dotted line in
Fig.~\ref{Fig1}, (a) of this paper. Numerical $f^{(3)}_c(\gamma_0)$ is shown 
by full black line with diamond symbols in this figure. 
It increases monotonously with $\gamma_0$ and
$f^{(3)}=f^{(1)}$ at $\gamma_0=\gamma_0^{(c)}$. 

How does hydrodynamic memory affect the Risken's phase diagram? We answer first this
important question. Profound memory effects for $\rho=\rho_f$ (neutrally buoyant condition) and
$\rho=\rho_f/2$ (a relatively lighter than fluid particle) used mostly in the
numerical simulations below profoundly change the phase diagram in Fig.~\ref{Fig1},
where also the limiting case of $\rho\to 0$ (maximal memory effects) is depicted. In
all three cases, numerical data are well approximated by a stretched-exponential
dependence $f_c^{(3)}\approx 1-\exp[-(\gamma_0/b)^a]$, with $0<a<1$, and $b>0$ shown
in the plot by dashed indigo line ($\rho=\rho_f$), full blue 
line ($\rho=\rho_f/2$), and dash-dotted green line ($\rho\to 0$), which come through the corresponding
different symbols depicting the numerical results. For $\gamma_0\ll b$ this yields 
$f_c^{(3)}\propto \gamma_0^a$ instead of
$f_c^{(3)}\propto \gamma_0$ in the memoryless case. It presents the first important
result of this paper. This dramatic change means: (1) For $\gamma_0$ smaller than
about $\gamma_0=0.5$ shown by the vertical line in panel (a) of Fig.
\ref{Fig1}, ever-increasing with lowering $\rho/\rho_f$ memory effects make an
effective friction larger. However, for $\gamma_0\gtrsim 0.5$, the opposite tendency
is seen in  panel (a). In particular, even for $\gamma_0$ essentially larger
than $\gamma_0^{(c)}$, the bistability region extends dramatically, cf.  panel
(b). It means that even for $\gamma_0>\gamma_0^{(c)}$ the memory correlations
can induce running solutions at $T=0$ and $f_c^{(3)}<f<f_c^{(1)}$, where, otherwise,
all Brownian particles would remain asymptotically trapped forever. Then, the memory
makes an effective friction smaller. This result agrees with the conclusions in Ref.
\cite{Seyler19} that hydrodynamic memory can induce transport at $T=0$ in a situation
where it would be absent otherwise, which was obtained therein for a critically
tilted piecewise linear potential. In the studied case, pertinent tilts can be
essentially smaller than the critical value $f_c^{(1)}$.  One should emphasize that
we are dealing here with a strongly driven transport.  It is the second result of
paramount importance, which provides a key for understanding our numerical results
below. 


\begin{figure*} 
\centering 
\includegraphics[width=0.47\textwidth]{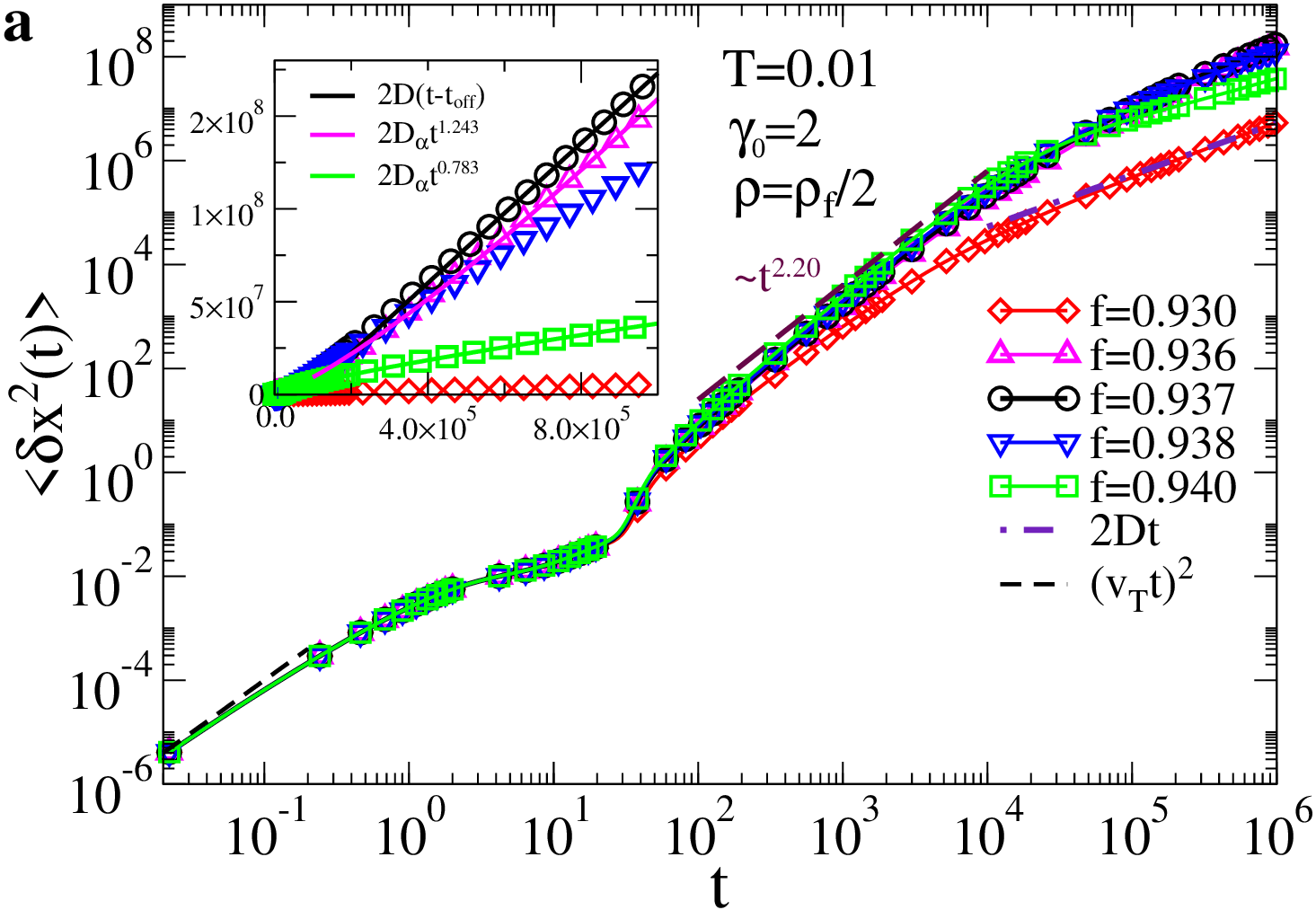}  \hspace{0.2cm}
\includegraphics[width=0.47\textwidth]{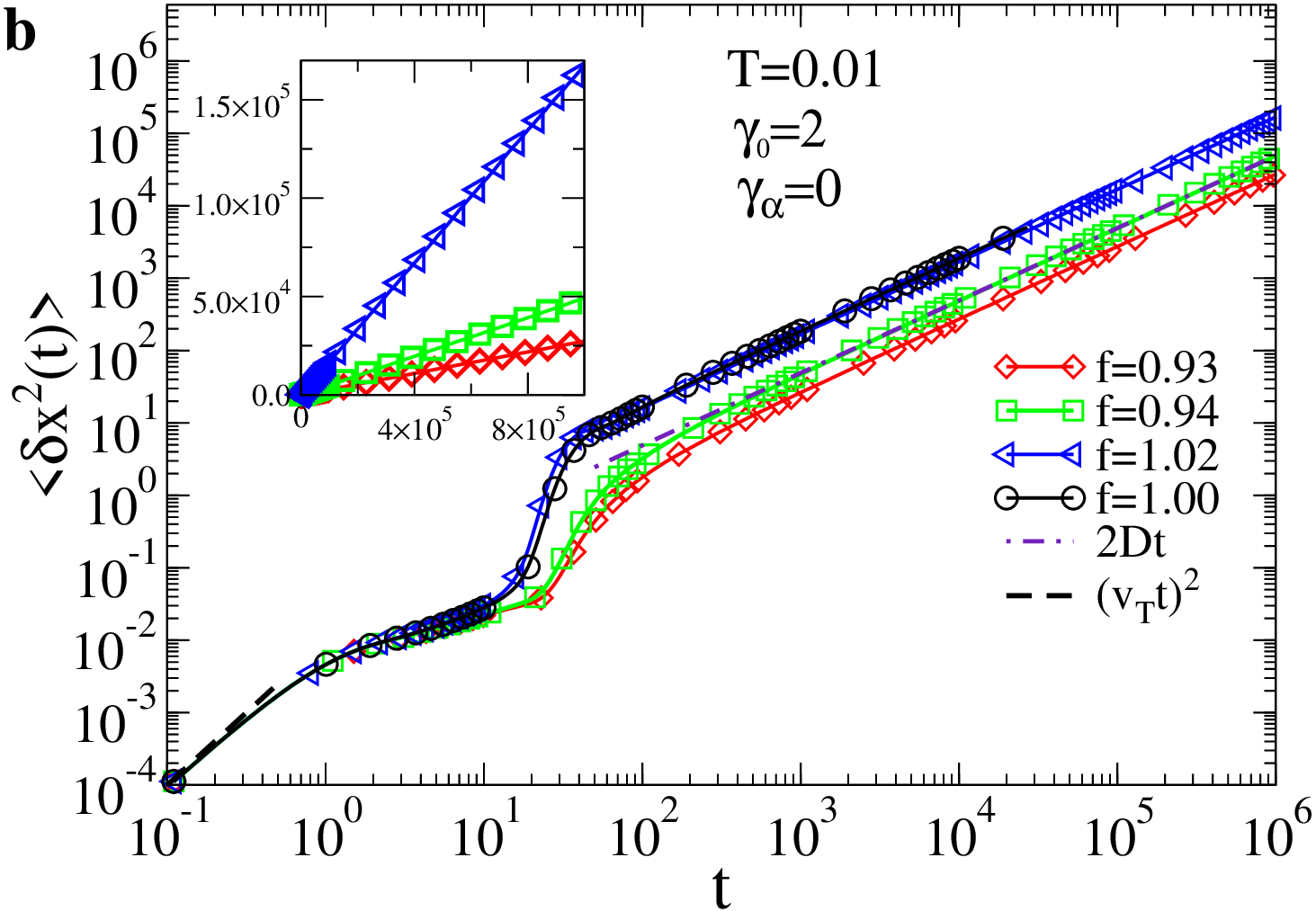} 
\caption{Dependence of the particles position variance on time for  (a)
non-Markovian and (b) Markovian diffusion at $T=0.01$,  $\gamma_0=2$  and
several values of force shown in the plot. In the non-Markovian case
$\gamma_\alpha=3$, which corresponds to $\rho=\rho_f/2$. $M=10^5$ particles are
used in the ensemble averaging. Particles are always initially localized at
$x=0$ within one potential well with their velocities Maxwell-distributed.
Initial diffusional spread is always briefly ballistic ($t\to 0$),
as indicated by the corresponding dashed black lines in main
plots. Insets show the same dependencies in linear plot. Panel  (b) makes clear
that without memory effects a normal diffusion regime is quickly established
once particles leave the potential well. However, memory effects in panel (a)
introduce a very long intermediate hyperdiffusive regime. Moreover, diffusion is
still anomalously fast at the end of simulations for $f$ close to the critical
tilt $f_c^{(2)}\approx 0.937$, see full black line with circles in (a),
including inset, which corresponds to resonance-like value of $D/D_0$ in Fig.
\ref{Fig4}, (a), and also full magenta line with triangles up for $f=0.936$ in
(a).  It is fitted by a superdiffusion dependence in the inset of (a). The case
of $f=0.937$ in this inset we fit with $2D(t-t_{\rm off})$ in the last half time
decade of simulations to derive $D$ from numerics,  see the main text. For
$f=0.940$ a very long subdiffusive regime astoundingly emerges in the last two
time decades of simulations, cf. the light green line with squares  in (a),
including inset. Full lines with different symbols correspond to different
values of $f$ shown in the main plots, except for the inset in (a), where three
lines correspond to the fits shown in this inset. } 
\label{Fig3}   
\end{figure*}

\subsection{Influence of memory on normal diffusion in unbiased potential}

Next, we expect that the dependence of the normal diffusion coefficient on $\gamma_0$ will also
be dramatically changed in a periodic potential, and we will check this hypothesis. Indeed, in
the case of unbiased diffusion, it is well-known that for strong friction, $\gamma_0\gg 1$,
$D\propto 1/\gamma_0$, and for the considered potential simple and well-known Lifson-Jackson
result $D=D_0/I_0^2(U_0/k_BT)$ holds \cite{LifsonJackson, RiskenBook,HanggiRevModPhys}. Here, $I_0(x)$ is a
modified Bessel function. This inverse friction proportionality holds at any $k_BT/U_0$. Less
known is that this scaling,  $D\propto 1/\gamma_0$, is valid also for weak friction, $\gamma_0\ll 1$, however, in the limit of large
barriers $U_0\gg k_BT$ only \cite{RiskenBook}. In the memoryless case, our numerics 
(depicted by black dashed line with diamonds) perfectly
agree with the Lifson-Jackson result for $\gamma_0 \geq 1$, see 
dash-dotted red line in Fig.~\ref{Fig2}. Hydrodynamic
memory, however, remarkable changes this result even for strong friction in the range
$2<\gamma_0<40$.  Namely, instead of the inverse friction dependence, our numerics are more
consistent with $D\propto 1/\gamma_0^{0.878}$, even for rather strong friction in the mentioned
range, see full blue line with circles and triangles in this figure. 
For much larger friction, the Lifson-Jackson result remains, however, valid, even in the
limiting case $\rho \to 0$. Furthermore, for a small friction $\gamma_0 < 1$, $D\propto
1/\gamma_0^{a}$ in Fig. \ref{Fig2}, with  $a\approx 0.413 $ for $\rho=\rho_f$ 
(dashed indigo line coming through triangles), and $a\approx
0.392 $ for $\rho=\rho_f/2$ (full blue line coming through circles), 
which is very different from $a\approx 0.83$ in the case of normal
diffusion (full black line coming through diamonds). 
The latter one deviates from the high-barrier theoretical value $a=1$ because
$k_BT/U_0$ is not small enough \cite{RiskenBook}, only $0.5$. The discovered non-trivial
dependencies of $D$ on $\gamma_0$ due to the memory effects present the third important result
of this work. It should be mentioned also that in the case of such equilibrium unbiased diffusion hydrodynamic memory always increases an effective friction because it makes the diffusion coefficient in Fig. \ref{Fig2}  smaller at all $\gamma_0$.

\begin{figure*}
\centering
	\includegraphics[width=0.47\textwidth]{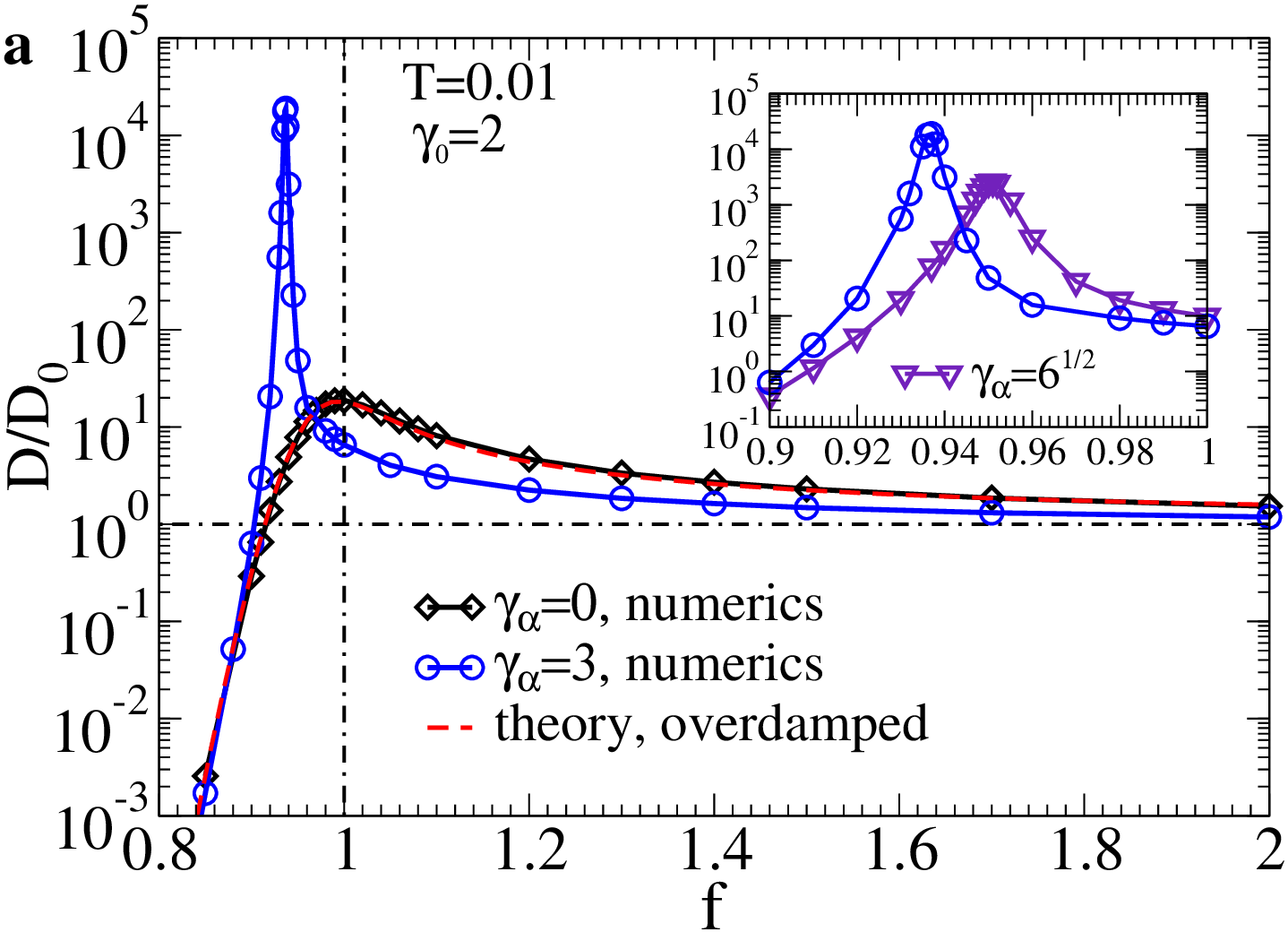} \hspace{0.2cm}
	\includegraphics[width=0.47\textwidth]{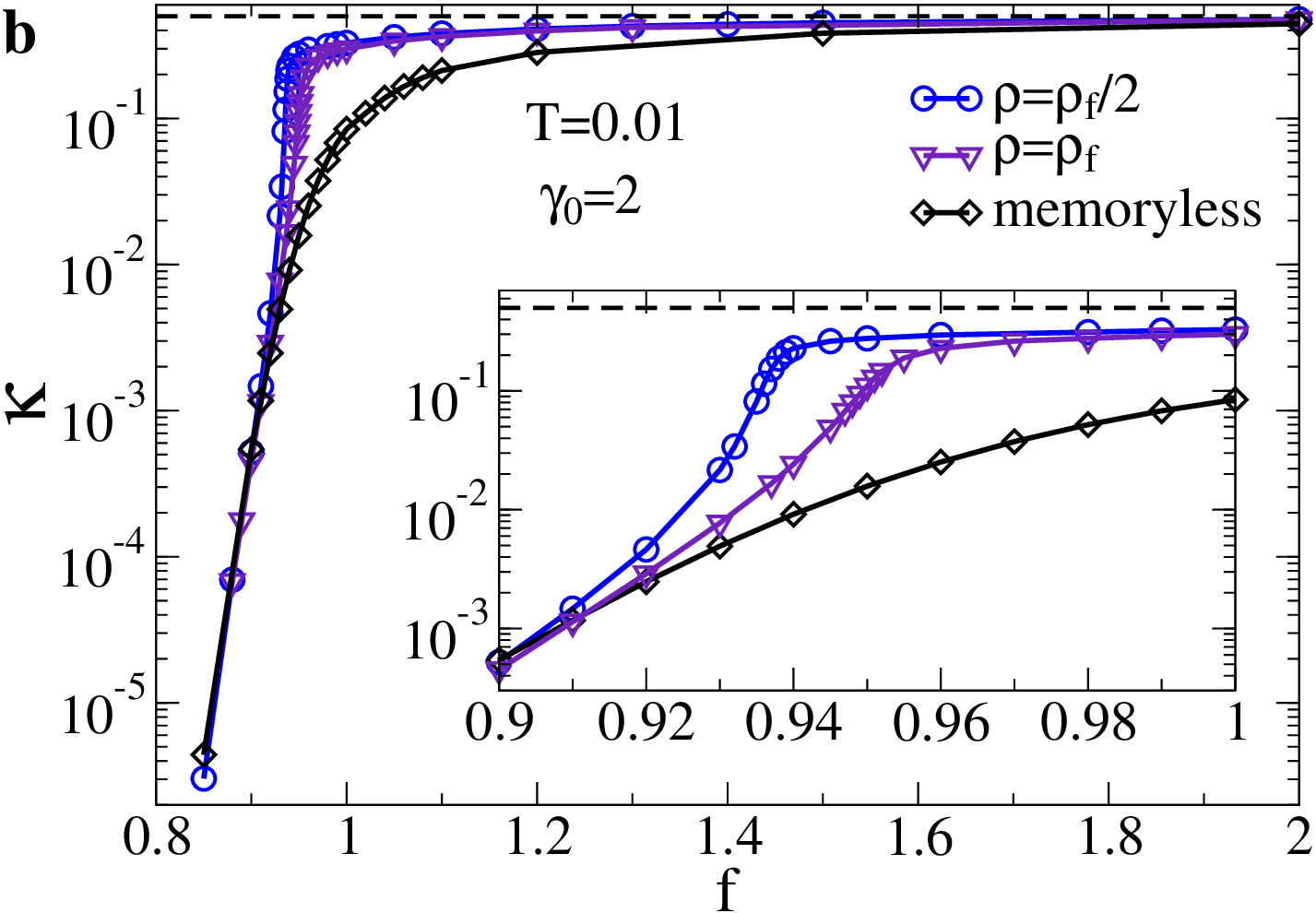} \\ \vspace{1cm}
	\includegraphics[width=0.47\textwidth]{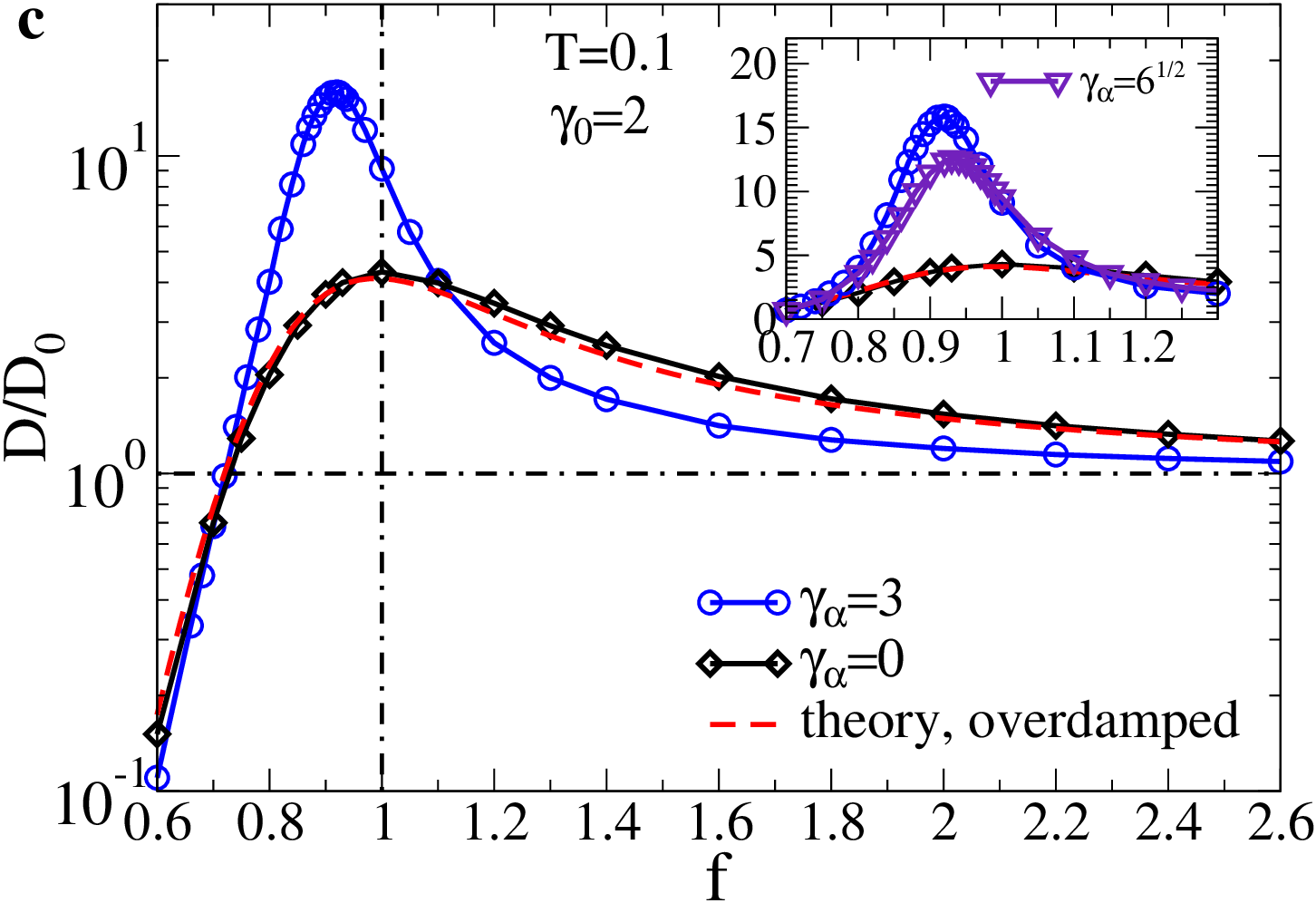} \hspace{0.2cm}
	\includegraphics[width=0.47\textwidth]{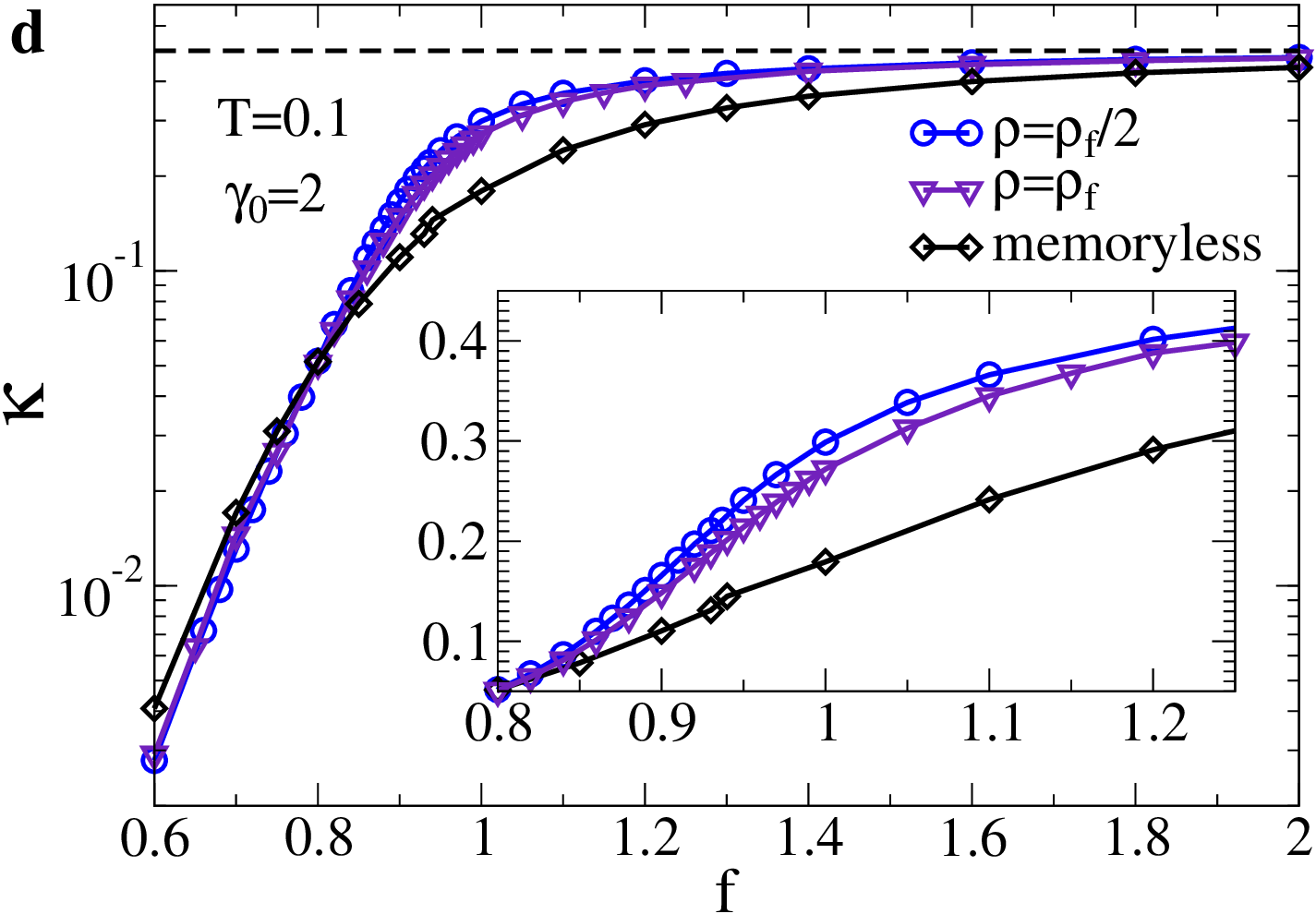} 	
	\caption{Diffusion coefficient and nonlinear mobility. (a,c) Enhancement of driven
diffusion over the  free-diffusion limit depending on the applied force $f$ for (a) $T=0.01$ and
(c)  $T=0.1$,
$\gamma_0=2$ and three values of $\gamma_\alpha$ shown in the plot. $\gamma_\alpha=3$
corresponds to $\rho=\rho_f/2$. The inset resolves a sharp peak for this value in the main plot
for a shorter range of $f$ and depicts for comparison the results for $\gamma_\alpha=\sqrt{6}$,
which correspond to $\rho=\rho_f$. The red dashed line in the main plots depicts the analytical
result of overdamped theory given by Eq. (8) in Ref. \cite{ReimannPRL01}. It remarkably agrees
with the memoryless result of simulations, which includes, however, the inertial effects
completely. (b,d) Nonlinear mobility $\kappa$ \textit{vs.} driving force $f$ for (b) $T=0.01$ and
(d)  $T=0.1$,
$\gamma_0=2$ and two values of $\gamma_\alpha$ corresponding to $\rho=\rho_f/2$ and
$\rho=\rho_f$, as well as for the memoryless case. The inset helps to resolve dramatic changes
around the corresponding critical values $f_c^{(2)}$, where the increase of mobility due to the
memory effects is dramatic. Both far below and far above $f_c^{(2)}$ the memory-induced effects
in $\kappa$ are negligible. For $f=2$, the linear mobility regime of $\kappa_0=1/\eta_0$ is
already almost achieved.	 } 
\label{Fig4}       
\end{figure*}

\subsection{Enormous boost of diffusion acceleration due to the memory effects}

Furthermore, we study an enormous acceleration of diffusion due to the memory effects in the
onset of an overdamped regime, $\gamma_0=2$, where the memoryless diffusion enhancement is
already nicely described by the results of the overdamped theory, cf. Eq. (8) in Ref.
\cite{ReimannPRL01}, see the corresponding comparison in our Fig. \ref{Fig4}, (a). It is
one of the greatest surprises of this work, which extends and complements recent findings in
Ref. \cite{GoychukPRL19}. For $\gamma_0=2$ and $\rho=\rho_f/2$, $f_c^{(3)}\approx 0.920$ in Fig.
\ref{Fig1}. The giant enhancement of diffusion is hence to expect for some critical value
$f_c^{(2)}$ such that $f_c^{(3)}<f_c^{(2)}<f_c^{(1)}$. For a small $\gamma_0$, this $f_c^{(2)}$
corresponds to the case, where the probabilities of trapped and running states become roughly
equal \cite{Marchenko14, LindnerSokolovPRE, GoychukPRL19}. At odds with intuition based on our
earlier results for $\gamma_0=0.1$ in Ref. \cite{GoychukPRL19}, this resonance-like enhancement
occurs for $\gamma_0$ essentially exceeding $\gamma_0^{(c)}$ of the Markovian case! Moreover,
now it does not correspond to the situation of equal probabilities of trapped and running
states, see below. The physics of this enhancement is hence different. Some of the results on
non-Markovian diffusion for $\gamma_0=2$, $\rho=\rho_f/2$ and $T=0.01$ are shown in Fig.
\ref{Fig3}, (a). Let us to  compare them with the results on the matching Markovian
diffusion in Fig. \ref{Fig3}, (b). First of all, in the Markovian case the normal
diffusion regime is already well-established on the scale exceeding the lattice period $2\pi$, 
$\langle \delta x^2(t)\rangle>4\pi^2$, for all values of $f$ in this plot. Quite on the
contrary, a very long regime of transient superdiffusion emerges in panel (a), which
extends on huge many potential periods, when $f$ becomes close to the resonance-like value
$f_c^{(2)}\approx 0.937$  (for $\rho=\rho_f/2$), see in Fig. \ref{Fig4}, (a). To derive
the results for the asymptotic value $D/D_0$ from the numerical data in Fig. \ref{Fig3}, we fit
the $\langle \delta x^2(t) \rangle $ dependence by $2D(t-t_{\rm off})$, where $t_{\rm off}$ is
some offset time required to account for a very long transient period of anomalous diffusion. It
can be neglected only for $Dt\gg Dt_{\rm off}$ while deriving $D/D_0$ from numerics.  For
example, for a sub-resonance value $f=0.93< f_c^{(2)} $ in Fig. \ref{Fig3}, (a), the
asymptotic normal diffusion regime is already well established. However, for the resonance
value  $f=f_c^{(2)}=0.937$, it is not, see results depicted with circles
in the inset of Fig. \ref{Fig3}, (a), 
and compare with the case $f=0.936$ (triangles up) to
realize why one needs $t_{\rm off}$. The corresponding value of $D/D_0$ in Fig. \ref{Fig4},
(a) is an estimate from below. The fantastic thousandfold enhancement $D/D_0$ by about
$18944$ times at peak, over the result in neglect of the memory effects, which enhancement
factor is about $18.2$ times ``only'', is, in fact, even larger. However, we cannot quantify it better because the proper normal diffusion limit is not reachable in simulations. With diminishing memory effects
along with increasing $\rho/\rho_f$, this enhancement weakens, see inset in Fig. \ref{Fig4},
(a), for neutrally buoyant case $\rho=\rho_f$ with $\gamma_\alpha=\sqrt{6}$ at $\gamma_0=2$, and,
nevertheless, it is still impressively strong. In this case, see inset in Fig. \ref{Fig4},
(a), $f_c^{(2)}\approx 0.951$ with $D/D_0\approx 2590$ at maximum. Generally, with
diminishing $\gamma_\alpha$ at fixed $\gamma_0> \gamma_0^{(c)}$, $f_c^{(2)}$ moves towards
$f_c^{(1)}$ and $D$ diminishes gradually to its memoryless value.

Next, at $f=0.936$ and $f=0.937$ in Fig. \ref{Fig3}, (a), diffusion is anomalously fast,
even near to the end simulation. Inset therein makes this clear for $f=0.936$, where $\langle
\delta x^2(t)\rangle\propto t^{\kappa_d}$ with $\kappa_d=1.243$, which is a fit alternative to $2D(t-t_{\rm
off})$ dependence that is not shown for this $f$ value. It is a striking result: Even for
$\gamma_0=2$, hydrodynamic memory effects can turn normal diffusion into superdiffusion over
many time decades corresponding to thousands of lattice periods! This long-lasting transient
superdiffusion is explained by kinetic heating, as Fig.~\ref{Fig6} of Appendix C makes
clear. Indeed, the kinetic temperature $T_k(t)=m^*\langle \delta v^2(t) \rangle/k_B $
 \cite{Brilliantov04, SieglePRL10, SiegleEPL11, Marchenko12} defined by the the variance of the
velocity distribution $P(v)$ increases dramatically, i.e., Brownian particles become kinetically
hot, with their $T_k$ substantially exceeding $T$ of the surrounding fluid. Velocity
distribution in Appendix C is also not Maxwellian, bimodal in the regime of enhanced
diffusion. However, differently from the case of small $\gamma_0$ \cite{Marchenko14,
LindnerSokolovPRE, GoychukPRL19}, for the considered large $\gamma_0$, $f_c^{(2)}$ does not
correspond, even approximately, to the situation where the local maximum of $P(v)$, which corresponds to the running
state, compares in amplitude with the local maximum corresponding to the trapped states. 
Moreover, for $f=f_c^{(2)}>f_c^{(3)}$, there are actually two running substates (within a
two-state approximation), as $P(v)$ in  Fig.~\ref{Fig7} reveals. Its first local maximum does not
correspond to $v=0$. Moreover, the integral $\int_{-\infty}^{v_c}P(v)dv$, where $v_c$ is the
velocity value, which separates two running substates, indicates that nearly 90\% of all
particles belong to the first running substate. It is very different from the case of low
friction. In the latter case, the peak of $D/D_0$ does correspond to the situation, where,
roughly speaking, one-half of the particles are temporarily trapped, while another one-half
run \cite{Marchenko14, LindnerSokolovPRE},  also in the presence of memory
effects \cite{GoychukPRL19}. Hence, for a sufficiently large $\gamma_0$, the mechanism of a huge
enhancement due to memory effects differs from one established for small
$\gamma_0$ \cite{GoychukPRL19}. Nevertheless, for strong memory
effects, $f_c^{(2)}$ seems to still roughly correspond to the maximum of stationary
non-equilibrium $T_k^{\rm (st)}$ \textit{vs.} $f$, like for a small
$\gamma_0$ \cite{GoychukPRL19}, as we detail in Appendix C. 

It is very different from the
matching case, where the memory effects are neglected.
Indeed, for the considered $\gamma_0$ and in the memoryless case,  the maximum of $D$ has nothing
in common with the maximum of $T_k^{\rm (st)}$, cf. Appendix C. The particles can become
kinetically very hot also in the absence of memory effects at $\gamma_0=2$, see in Fig. \ref{Fig6},
(b). It seems to be first in contradiction with the fact that the overdamped theory
remarkably well describes the numerical results on $D/D_0$ enhancement in our Fig. \ref{Fig4}, (a). 
We defer a detailed explanation of this puzzle somewhere else. In short, due to
inertial effects, there emerge very fast oscillations in the non-equilibrium but stationary VACF
for $f>1$, even for sufficiently large $\gamma_0\sim 2-10$. Following Green-Kubo relation \cite{KuboBook}, diffusion coefficient is integral of
VACF and those fast oscillations reduce $D$ in spite of still growing (for $f>1$, in the memoryless
case) $T_k^{\rm (st)}$. It resolves the discussed apparent contradiction. However, it leads to a
paradox: Becoming ever hotter particles (for a certain intermediate interval of overcritical forces
$f>1$) diffuse ever slower, and $D/D_0$ drops for $f>1$. This very striking non-equilibrium
phenomenon emerging due to inertial effects in nonlinear \textit{memoryless} driven dynamics for a
sufficiently large, but not too large friction $\gamma_0$ was entirely overlooked thus far. It will
be studied in detail somewhere else. The inertial effects in the Brownian motion are highly
nontrivial and remain insufficiently studied until now, even in the simplest paradigmatic systems
like one considered.   

Next, quite embarrassing, for $f=0.94$, a very prolong hyperdiffusive, faster than ballistic,
regime (see the main plot in Fig. \ref{Fig3}, a) changes in the last time decade into a
subdiffusive regime with $\langle \delta x^2(t)\rangle\propto t^{0.783}$ (see 
light green line with squares in inset therein). The
latter one is, for sure, transient. Nevertheless, its appearance for an already saturated with time
$T_{\rm k}^{(\rm st)}$, which exceeds $T$ by more than four and a half times, see in Fig.~ \ref{Fig6} is
physically really puzzling. Mathematically, it, of course, just corresponds to a very long
transition from superlinear $\langle \delta x^2(t)\rangle\propto t^{2.2}$ scaling with time to an
asymptotically linear one. Quite paradoxically, we are dealing here with a hot subdiffusion. For a
small $\gamma_0$, such a regime occurs due to a transient cooling after the maximum of $T_k(t)$ in
time is passed \cite{GoychukPRL19}. It was revealed also for a periodically driven memoryless underdamped diffusion in a ratchet potential \cite{Spiechowicz16}.  In the present case, the underlying mechanism is, however, different. Notice also that
in this case $t_{\rm off}$ is negative and $2D(t-t_{\rm off})$ fit overestimates $D$. With a
further increase of $f$, the asymptotically normal regime is gradually established.

\subsubsection{Diffusion at higher temperature}

With the increase of temperature $T$, the influence of memory effects on the diffusion 
enhancement becomes smaller. Nevertheless, for $T=0.1$ in Fig. \ref{Fig4}, (c) it is
still manifestly present. First, the diffusion maximum occurs at some $f_c^{(2)}<1$. Second, at its maximum the
enhancement is three to four times stronger than  in the memoryless case, that also is pretty well
described by the analytical result of the overdamped theory.  Indeed, for $\rho=\rho_f/2$ at in
Fig. \ref{Fig4}, (c),  $f_c^{(2)}=0.92$. Notice that it equals $f_c^{(3)}$ in this case and
$D/D_0\approx 15.87$ at maximum. It should be compared with the maximum $D/D_0\approx 4.32$ at
$f=1$ in the memoryless case. The memory-induced boost is about by $3.67$ times. For $\rho=\rho_f$
therein, $f_c^{(2)}\approx 0.93$ with maximal $D/D_0\approx 12.57$. The increase is still
impressive nearly $2.91$ times. The effect is not small.

\begin{figure*}
	\includegraphics[width=0.47\textwidth]{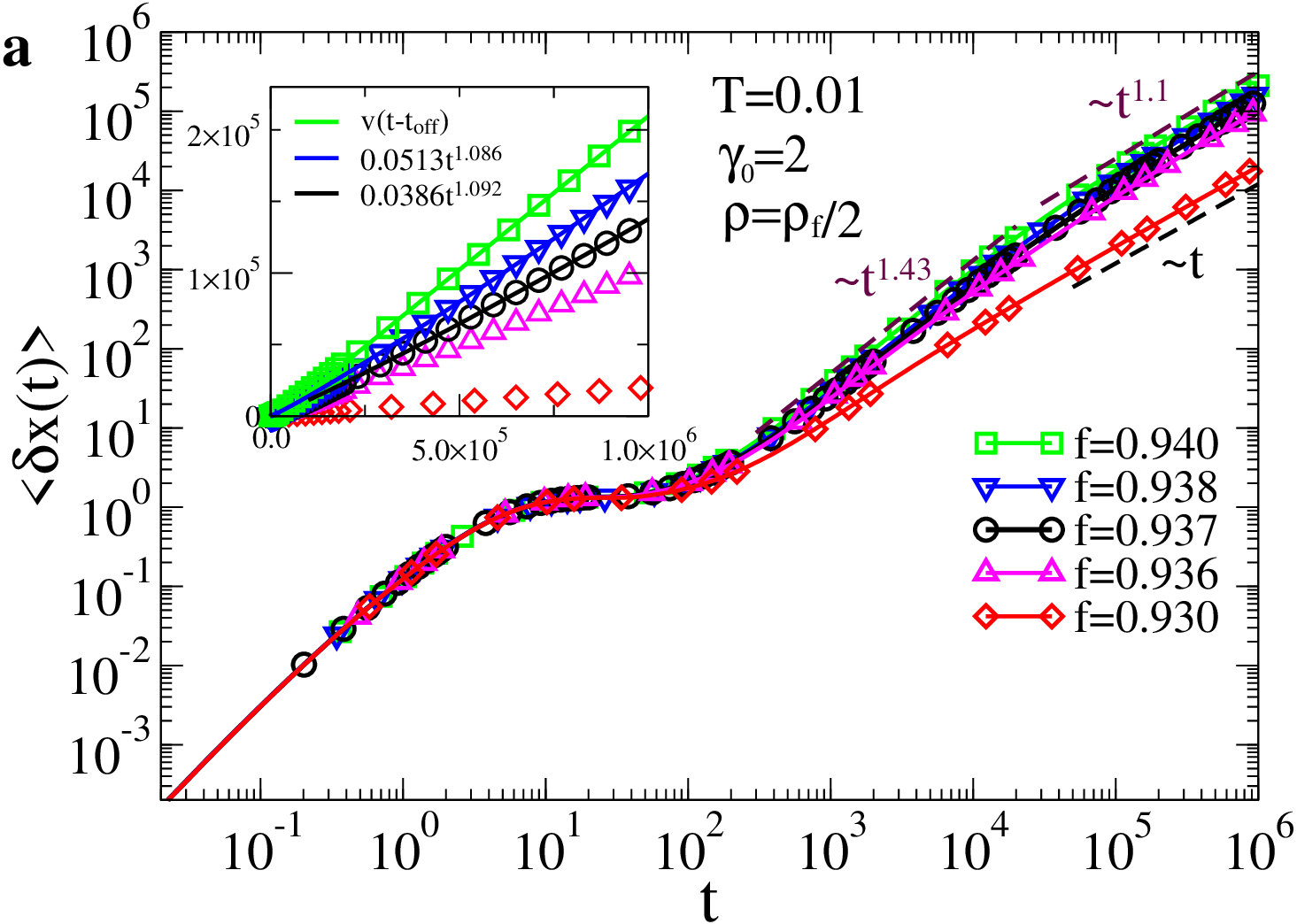} 
	\hspace{0.4cm}	
	\includegraphics[width=0.47\textwidth]{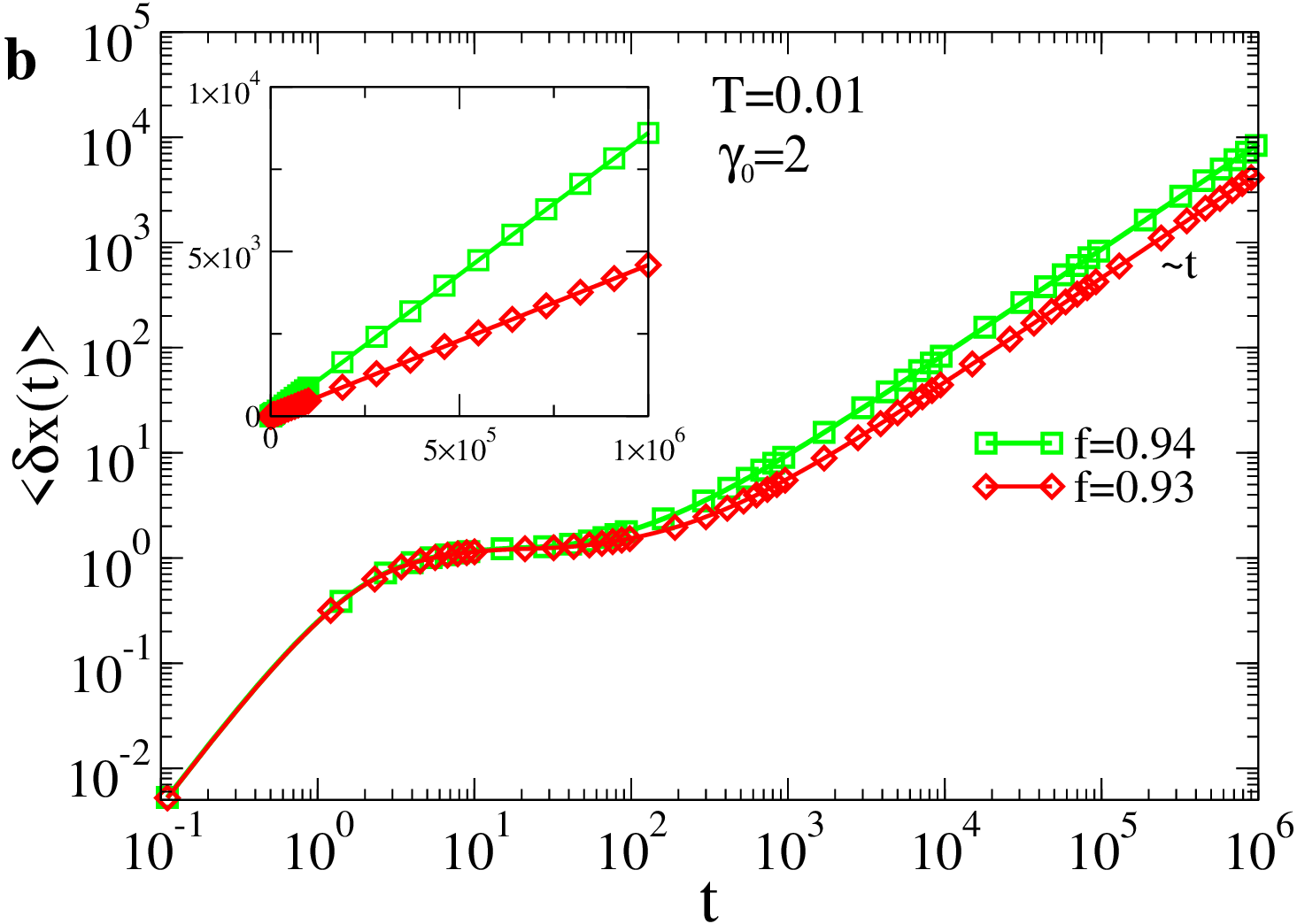}
	
	\caption{Dependence of the particles displacement on time in (a) non-Markovian and
	(b) Markovian transport at $T=0.01$,  $\gamma_0=2$  and several values of force shown
	in the plot. 
	In panel (a), $\gamma_\alpha=3$, which corresponds to $\rho=\rho_f/2$.
	$M=10^5$ particles are used in the ensemble averaging. Particles are always initially
	localized at $x=0$ within one potential well with their velocities Maxwell-distributed. Insets
	show the same dependencies in linear plot. Panel (b) makes clear that without memory
	effects a normal transport regime is quickly established once particles leave the potential
	well. However, memory effects in panel (a) introduce a very long intermediate
	supertransport regime. Moreover, transport is still anomalously fast at the end of simulations
	for $f$ close to the critical tilt $f_c^{(2)}\approx 0.937$, see inset,  which corresponds to
	resonance-like value of $D/D_0$ in Fig. 4, (a) of the main text.  
	Full lines with different symbols correspond to
matching values of $f$ shown in the main plots, except for the inset in (a), where three lines
correspond to the fits shown in this inset. }

	\label{Fig5}      
\end{figure*}

\subsection{Hydrodynamic memory boosts nonlinear mobility in a subcritical tilt region}

Finally, we study the influence of memory effects on the particles' nonlinear mobility 
$\kappa(f)=v(f)/f$, where $v$ is the mean particles velocity $v=\langle \delta x(t)\rangle/t$
defined at the last half time decade of simulations by using a $v(t-t_{\rm off})$ fit to
numerical $\langle \delta x(t)\rangle$.
The mean displacement of particles is shown in Fig. \ref{Fig5}, (a), 
for the case with strong memory, $\rho=\rho_f/2$, $T=0.01$, $\gamma_0=2$
and several values of force in a narrow interval, $f=0.93-0.94$, around $f_c^{(2)}=0.937$.
It should be compared with Fig. \ref{Fig5}, (b), where the matching
memoryless case, $\gamma_\alpha=0$, is depicted. First of all, it is worth noting
that in the limit $t\to 0$, the initial transport is universally ballistic, $\langle
\delta x(t)\rangle\sim ft^2/2$ ($m^*=1$). For a very brief initial time period, the
periodic potential does not matter. The particles are prepared at $x=0$, which is not
a mechanically equilibrium state in the biased case, and they move first accelerating
towards nearest potential minimum. Then, they start to equilibrate and $\langle
\delta x(t)\rangle$ temporary saturates. Notice also that the range of $f$ variation
in this figure is so small that all the curves practically coincide during the
equilibration process with the potential well. Next, the particles start to escape
out of the potential well and travel over many potential periods being driven by $f$;
become occasionally trapped in other potential wells during this process and
rereleased (notice that $f$ is rather close to $f_c^{(1)}=1$ from below). In the
memoryless case, transport is practically normal, $\langle \delta x(t)\rangle \propto
v(t-t_{\rm off})$ (where $t_{\rm off}$ is some offset time) once $\langle \delta
x(t)\rangle$ exceeds $2\pi$ -- the potential period. Also, in the case with memory,
but for subcritical $f=0.930$ shown by red line with diamonds 
in panel (a), the normal transport regime is
established relatively fast. However, for $f$ closer to $f_c^{(2)}$, a dramatic
enhancement of transport occurs. Notice that then transport becomes very sensitive to
tiny $f$ variations (compare with the initial regime for $t<100$!). Remarkably, a
very long transient supertransport regime $\langle \delta x(t)\rangle \propto
t^{\kappa_t}$ with $\kappa_t\approx 1.43$ emerges, which can cover about one thousand
potential periods. Of course, even in this regime, the transport is slower in
absolute terms than in the absence of hindering periodic potential. However,
hydrodynamic memory greatly accelerates transport in the periodic potential
near-to-critically tilted, i.e., the memory effects in synergy with thermal fluctuations greatly help to overcome the
residual potential bumps on the way. Interesting, even at the end of simulations
(which take several days being run with double numerical precision and $M=10^5$
particles in parallel on high-performance professional GPU processors, for one curve
presented) the transport remains anomalously fast, $\kappa_t\approx 1.1$, in this case,
see inset of panel (a) for more detail. The fit with $v(t-t_{\rm off})$, see,
e.g., in the discussed inset for $f=0.940$ then still underestimates the actual value
of $v$. In Fig. 4, (b) of the main text, we provided the nonlinear mobility
based on such an estimate in the pertinent cases, where the normal transport regime
was not possible to reach in numerics.

The results for mobility are depicted at $T=0.01$ in Fig.
\ref{Fig4}, (b)   for three values of $\gamma_\alpha$, including the memoryless case. For
$f$ smaller than $f=0.9<f_c^{(3)}$, the influence of memory is negligible in this plot.
Likewise, for a very large $f$, the regime of linear mobility in the absence of potential,
$\kappa_{0}=1/\eta_0=0.5$, is gradually achieved already for $f\geq 2$. However, near to
$f_c^{(2)}$ the enhancement of mobility by the memory effects is tremendous. For
$\rho=\rho_f/2$, at $f=0.94$, $\kappa=0.2288$, whereas without memory effects it is merely
$0.009156$. The boost of mobility is about 25 times! This is a very striking effect. Likewise,
for $\rho=\rho_f$, at $f=0.96$, $\kappa=0.2283$, whereas without memory effects it is merely
$0.02524$. The memory-caused increase is by impressive nearly 9 times. 

Important to mention is that for $T=0.01$ and $f$ close to $f_c^{(2)}$, the
transport is, in fact, anomalously fast during the major period of simulations,
as Fig. \ref{Fig5}, (a) reveals. First, $\langle \delta x(t)\rangle
\propto t^{1.43}$ for intermediate times after particles started to leave the
potential well and move in the force direction. This super-transport regime 
lasts for $t$ until about $2\times 10^4$. During this time, particles move over
about a thousand potential periods. Notice that the transport power-law exponent
$\kappa_t=1.43$ is close to $1.5$, which would correspond to the case of Stokes
friction contribution \textit{ad hoc} neglected \cite{SiegleEPL11}. One can
state that the corresponding super-transport regime is manifested here in the
presence of Stokes friction, which is an important result. Second, even at the
end of the simulation, the corresponding power-law exponent still did not relax
to unity being about $1.1$. The corresponding estimates of the mean velocity,
like one shown in the inset of Fig. \ref{Fig5}, (a) for $f=0.94$
underestimate, in fact, the corresponding value of $v$ and the mobility in Fig.
\ref{Fig4}, (b). Comparison with Fig. \ref{Fig5}, (b), where such
a regime is absent, makes clear that this supertransport emerges due to
hydrodynamic memory effects.

\subsubsection{Transport at higher temperature}

Finally, we provide the readers with the results on nonlinear mobility enhancement
for a larger $T=0.1$ or smaller $U_0/(k_BT)=10$ in  Fig. \ref{Fig4}, (d).  The
memory-induced increase of mobility becomes less impressive, and, nevertheless, it
remains still significant. For example, in this figure, at $f=0.93$, 
$\kappa=0.2106$ for $\rho=\rho_f/2$, $\kappa=0.1882$ for $\rho=\rho_f$, and
$\kappa=0.1310$ in the memoryless  case. The enhancement is by 60.75\% and 43.66\%
with respect to the memoryless case, correspondingly. It is not small at all.

We conclude that also at higher temperatures the memory effects can significantly
boost both diffusion and transport in near-to-critically tilted periodic potentials
over the case, where such effects are neglected. Hydrodynamic memory can also
suppress diffusion outside the resonance-like critical enhancement regime,  see,
e.g., in Fig. \ref{Fig4}, (a) for $f>1$ and in Fig. \ref{Fig4}, (c) for
$f>1.15$ and $f<0.7$.  Similar suppression was  earlier  discussed in Ref.
\cite{GoychukPRL19} for a weak friction case of $\gamma_0=0.1$.  Transport is
generally enhanced for $f>f_c^{(3)}$, however, it can also be suppressed for 
$f<f_c^{(3)}$, see, e.g.,  in Fig.  \ref{Fig4}, (d) for $f<0.8$. In this
respect, the case of moderately strong $\gamma_0=2$ considered here is different from
the case of small $\gamma_0=0.1$ investigated in Ref. \cite{GoychukPRL19}. Our Fig.
\ref{Fig1} explains why. In any case, influence of hydrodynamic memory on strongly
nonequilibrium stochastic transport can be very significant even for moderately
strong friction in the case of sufficiently light particles.

\section{Summary and conclusions}

In summary, in this paper, we showed within a paradigmatic model of driven nonlinear Brownian
transport and diffusion that hydrodynamic memory effects primarily neglected thus far in the
theory of nonlinear  Brownian motion could profoundly influence both diffusion and transport
even for a relatively strong Stokes friction. First, enormous resonance-like enhancement of
diffusion can occur for a potential tilt, which is subcritical for such a Stokes friction taken
alone, where the pertinent overdamped theory \cite{ReimannPRL01} describes already very well the
numerical results in the negligence of such memory effects. This memory-induced surplus
enhancement can be giant for light particles, by several orders of magnitude, depending on temperature and the ratio $\rho/\rho_f$ which measures strength of the memory effects. Second, also transport can be enhanced
enormously at the corresponding resonance-like tilt. Third,  the transient superdiffusive and
supertranport regimes can last for a long time while covering thousands of lattice periods.
Particles can become kinetically very hot in these anomalous regimes, with their kinetic
temperature well above the temperature of the surrounding liquid.   Fourth, even undriven
thermally equilibrium diffusion in periodic potentials exhibit novel features manifested by an
inverse fractional dependence of the diffusion coefficient on the Stokes friction strength.

The experimental verification of these intriguing and highly surprising effects can be expected,
cf. Appendix B, in least viscous liquids like liquid helium at $T=4$ K
(above $\lambda$ point, still a normal fluid), for hollow microparticles with tailored $\rho$, which are trapped in optically
created potentials \cite{LeeGrierPRL}, or even in more viscous yet more common fluids like
diethyl ether for nanoparticles in nanoimprinted periodic potentials created by methods of
lithography \cite{Guo}. In such micro- and nanofluidic systems, the inertial effects in Brownian
motion can become essential being greatly amplified by the hydrodynamic memory effect, as this
work showed. We expect that it will attract the interest of not only theoreticians but also
experimental scientists and spark subsequent research work.    

\section*{Acknowledgment} 

We acknowledge support by the Regional
Computer Centre Erlangen, Leibniz Supercomputing Centre of the Bavarian Academy of Sciences and
Humanities, as well as University of Potsdam (Germany), which kindly provided GPU high-performance
computational facilities for doing this work. This research was funded by 
the Deutsche Forschungsgemeinschaft (German Research Foundation),
Grant GO 2052/3-2.

\appendix

\section{Numerical approach}

Numerical integration of FLE (\ref{FLE1}) is based on an approximation of the power-law scaling part of
memory kernel by a sum of exponentials (a Prony series expansion) and hyper-dimensional Markovian embedding
of underlying non-Markovian dynamics \cite{GoychukPRE09, GoychukACP12,SiegleEPL11, GoychukPRL19}. The method
works very well and leads to results, which often practically coincide 
within the numerical precision
tolerance of 5-6\% (can be made better) with the analytical results available in case of linear dynamics.
The memory kernel approximation reads \cite{SiegleEPL11,GoychukPRL19}  
\begin{align}
\eta(t)= \sum^N_{i=1}\eta_i\Bigg[2\delta(t)
- \nu_i \exp\left(-\nu_i t\right)\Bigg]\;,
\label{fk}
\end{align}
where $\eta_i=k_i/\nu_i$, $k_i= C_\alpha(b)\eta_\alpha\nu_i^\alpha/|\Gamma(1-\alpha)|$,  and
$\nu_i=\nu_0/b^{i-1}$. The sum of exponentials obeys a fractal scaling with a scaling parameter $b$. It
approximates the power-law decay \cite{Hughes, PalmerPRL85, GoychukPRE09, GoychukACP12} of this memory
kernel, so that $\int_0^\infty \eta(t)dt=0$. The choice of $\nu_0$ is related to the time step of simulation
$\Delta t$, which was $\Delta t=0.002$ in most simulations. To avoid numerical instability, $\nu_0 \Delta t$
should be smaller than one. 

The power-law regime extends in this approximation from a short time (high-frequency) cutoff,
$\nu_0^{-1}$, to a large time  (small frequency) cutoff, $\tau_h=\tau_l b^{N-1}$. The choice of
$N$ is dictated by the maximal time $t_{\rm max}$ of simulations: $\tau_h$ should exceed $t_{\rm
max}$ by at least several times. The accuracy of the approximation between two cutoffs is
controlled by the scaling parameter $b>1$. The smaller $b$, the better the accuracy. However, a
larger $N$ is then required.  With $b=5$ and
$C_\alpha(b)=1.78167$ \cite{SiegleEPL11,GoychukPRL19}  it is about 6\% for $t$ between $0.05$ and
$10^6$ for $\nu_0=100$ and $N=13$. It can be slightly improved to 5\% by choosing
$C_\alpha(b)=1.816$, which is used in most simulations in this paper. When required, the
discussed accuracy can drastically be improved to about $0.003\%$ between $0.07$ and $10^6$ for
$\nu_0=100$ and $N=38$ with $b=2$ and $C_\alpha(b)=0.782134$. This choice would, however, also
essentially increase the simulation time because of a much larger embedding dimension. Since we
are interested in reaching a maximal time range in computer simulations, we use the same
embedding with $b=5$ and $N=13$ as earlier \cite{SiegleEPL11, GoychukPRL19}.  Even in this case,
simulations are very time-consuming. It takes several days to reach $t_{\rm max}=10^6$ on
professional GPU processors (double precision accuracy) required for the reason of a trivial
parallelization: $M=10^5$ independent Brownian particles  (trajectories) were propagated in
parallel for doing ensemble averaging. Sufficiently large $t_{\rm max}$ is required given very
long transient regimes.  For the studied problem, in the neglect of memory effects, it suffices
to use $\Delta t=0.01$ in numerics \cite{LindnerSokolovPRE} done here with the second-order
stochastic Heun algorithm \cite{GardBook}. In the presence of memory effects, five times smaller
$\Delta t=0.002$ was sufficient.  

For doing Markovian embedding, one introduces a set of $N$ auxiliary variables  $u_i$ such that 
the corresponding embedding dynamics in the hyperspace of dimension $D=N+2$ 
reads \cite{SiegleEPL11,GoychukPRL19} 
\begin{align}\label{emb}
\dot x(t)=&v(t)\nonumber \\
m^*\dot v(t)=&f(x,t)-\sum^N_{i=1}u_i(t)-(\eta_0+\eta_{\Sigma})
v(t) \nonumber \\
+&\xi_0(t)+\sqrt{2k_BT\eta_{\Sigma}}\zeta_0(t) \nonumber \\
\dot u_i(t)=&-k_i v(t)-\nu_iu_i(t)+\sqrt{2k_BT k_i\nu_i}\zeta_i(t) ,
\end{align}
for $i=1,...,N$, where $\eta_\Sigma=\sum_{i=1}^N \eta_i$. Furthermore, $\xi_0(t)$ and $\zeta_i(t)$, $i=1..N$, are $N+1$ 
delta-correlated in time and mutually uncorrelated white Gaussian noise sources of zero-mean and unit intensity, 
$\langle\zeta_i(t)\zeta_j(t^\prime)\rangle=\delta_{ij}\delta(t-t^\prime)$, for
$i,j=1..N$, $\langle \xi_0(t)\zeta_i(t')\rangle=0$.
However, the noise $\zeta_0(t)$ is chosen as a weighted, 
normalized sum of other $\zeta_i(t)$ \cite{SiegleEPL11,GoychukPRL19}, 
\begin{eqnarray}\label{corr}
\zeta_0(t)=\sum_{i=1}^N\sqrt{\frac{\eta_i}{\eta_\Sigma}}\zeta_i(t) \;.
\end{eqnarray}
The initial  $u_i(0)$ are sampled as independent Gaussian variables with zero mean and correlations   
$\langle u_i(0)u_j(0)\rangle=k_BT k_i\delta_{ij}$ \cite{GoychukPRE09,GoychukACP12,SiegleEPL11}.
Initially, particles were always prepared with their velocities Maxwell-distributed at temperature $T$
and localized sharply at $x=0$, which corresponds to the minimum of potential in the unbiased case
$f=0$. One assumes besides that $v=0$ for $t<t_0=0$ in Eq. (\ref{FLE1}). This is a non-equilibrium
initial preparation.

\section{Estimation of physical parameters.}

Here, we address physical systems, where the studied effects can be revealed experimentally. The crucial
issue here is a sufficiently small non-dimensional $\tilde \gamma_0$, which can be expressed as
\begin{eqnarray}
\tilde \gamma_0=6\sqrt{\frac{3\pi}{2}\frac{\rho_f}{2\rho+\rho_f}}\frac{x_0}{R}\theta,
\end{eqnarray}  
where $\theta=\sqrt{\frac{R\rho_f}{U_0}}\mu=\sqrt{\frac{R}{U_0\rho_f}}\zeta$, with
$\zeta=\rho_f\mu$ being the dynamic viscosity. The particles diameter should not be much larger
than the potential period $L=2\pi x_0$. For this reason, $x_0/R$ could hardly be much smaller
than $0.1$. For example, for Brownian particles in optical vortices in Ref. \cite{LeeGrierPRL},
$x_0=52.5$ nm and $R=740$ nm. This yields $x_0/R\approx 0.0709$. To arrive at smallest $\tilde
\gamma_0$, the strategy is hence to minimize $\theta$. The fluid with lowest known dynamic and
kinematic viscosity is liquid helium. At $T=4$ K, its dynamic viscosity is $\zeta=3.3\times
10^{-6}\;\rm Pa \cdot s$ \cite{Hands1986}. With density $\rho_f=125\;\rm kg/m^3$ this yields
$\mu=2.64\times 10^{-8}\;\rm m^2/s$. Let us take $U_0=100\;k_BT\approx 5.52\times 10^{-21}\;J$.
Then, $\theta\approx 3.417$, and for $\rho=\rho_f$, we obtain $\tilde \gamma_0\approx 1.823$,
which is a bit smaller than $\tilde \gamma_0=2$ used in this paper. The corresponding time units
in our simulations would be $\tau_0=1.26\times 10^{-5}$ s in physical units. Hence, the maximal
time $t_{\rm max}=10^6$ in our simulations would correspond to about 12.6 seconds. This is
interesting because the regimes of anomalous diffusion caused by hydrodynamic memory effects can
reach the time scale of seconds. The values $\tilde \gamma_0$ ten times smaller can be achieved
if to downscale $x_0$ and $R$ by a factor of hundred, i.e., for nanoparticles of $R\sim 7.4$ nm.
The corresponding periodic nanostructures to create, e.g., electrostatic periodic potentials for
charged Brownian particles can be produced by nanolithography \cite{Guo}, and nanosized hollow
particles with appropriate low mass densities  can also be tailored \cite{Bentz18}. Furthermore,
for more common fluids like diethyl ether at $T_r=298$ K, $\zeta=0.224\times 10^{-3}\rm Pa\cdot
s$ and $\rho=713.4\;\rm kg/m^3$. For $U_0=100\;k_BT_r\approx 4.114\times 10^{-19}\;\rm J$,
$R=740$ nm, $x_0=52.5$ nm we obtain $\theta\approx 11.247$, and for $\rho=\rho_f$, $\tilde
\gamma_0\approx 6$. Next, a 9-fold reduction of both $R$ and $x_0$ would yield $\tilde
\gamma_0\approx 2$. Hence, the study of inertial effects in Brownian motion, including
hydrodynamic memory effects, should be experimentally feasible. Some significant experimental
work in this direction was already done for parabolic potentials optically
created \cite{FranoschNature, HuangNatPhys, Kheifets14}. The case of periodic potentials is, however, more
challenging.

\section{Kinetic heating and velocity distribution}

\begin{figure*} \includegraphics[width=0.47\textwidth]{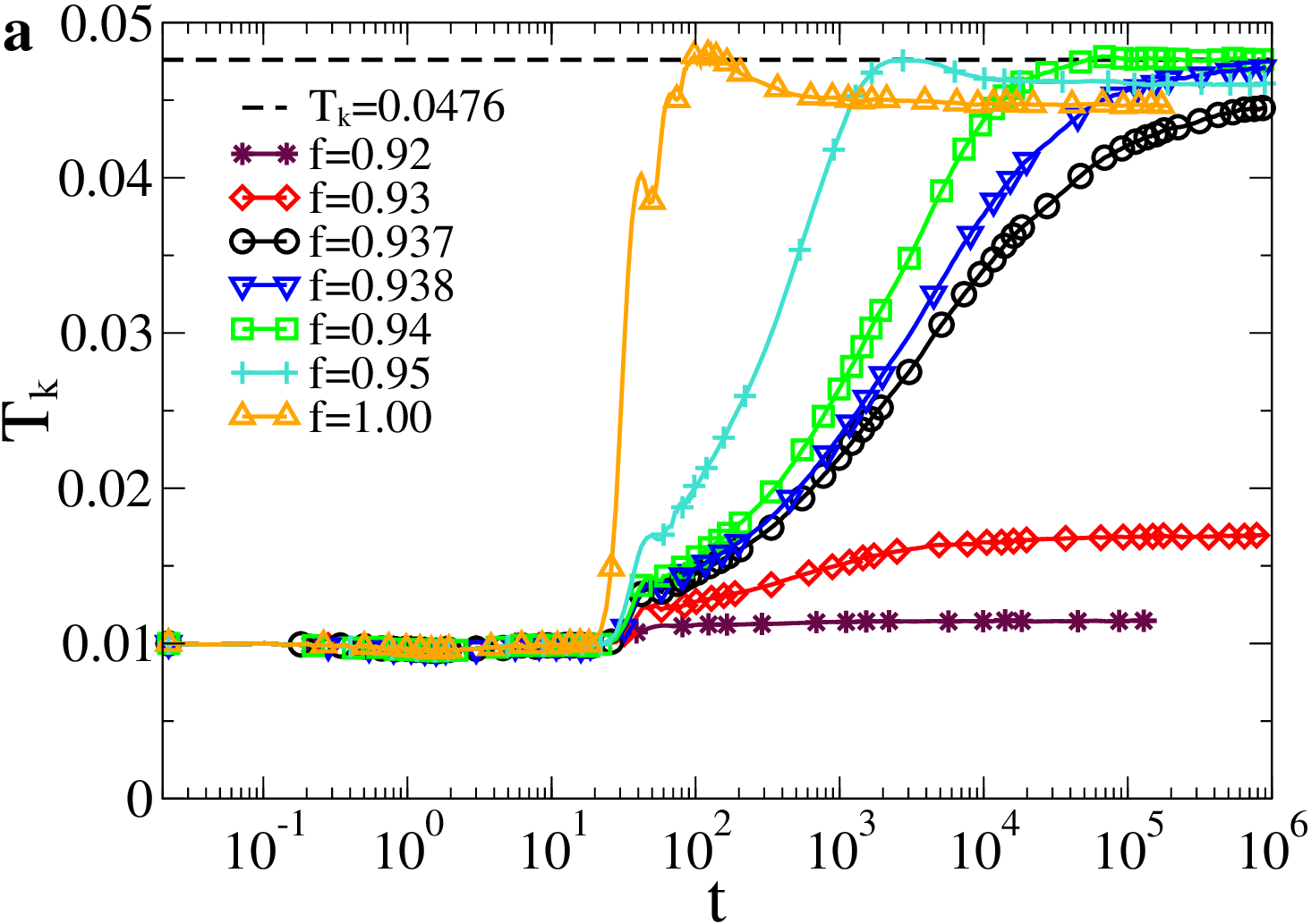} 
\hspace{0.4cm}	
\includegraphics[width=0.47\textwidth]{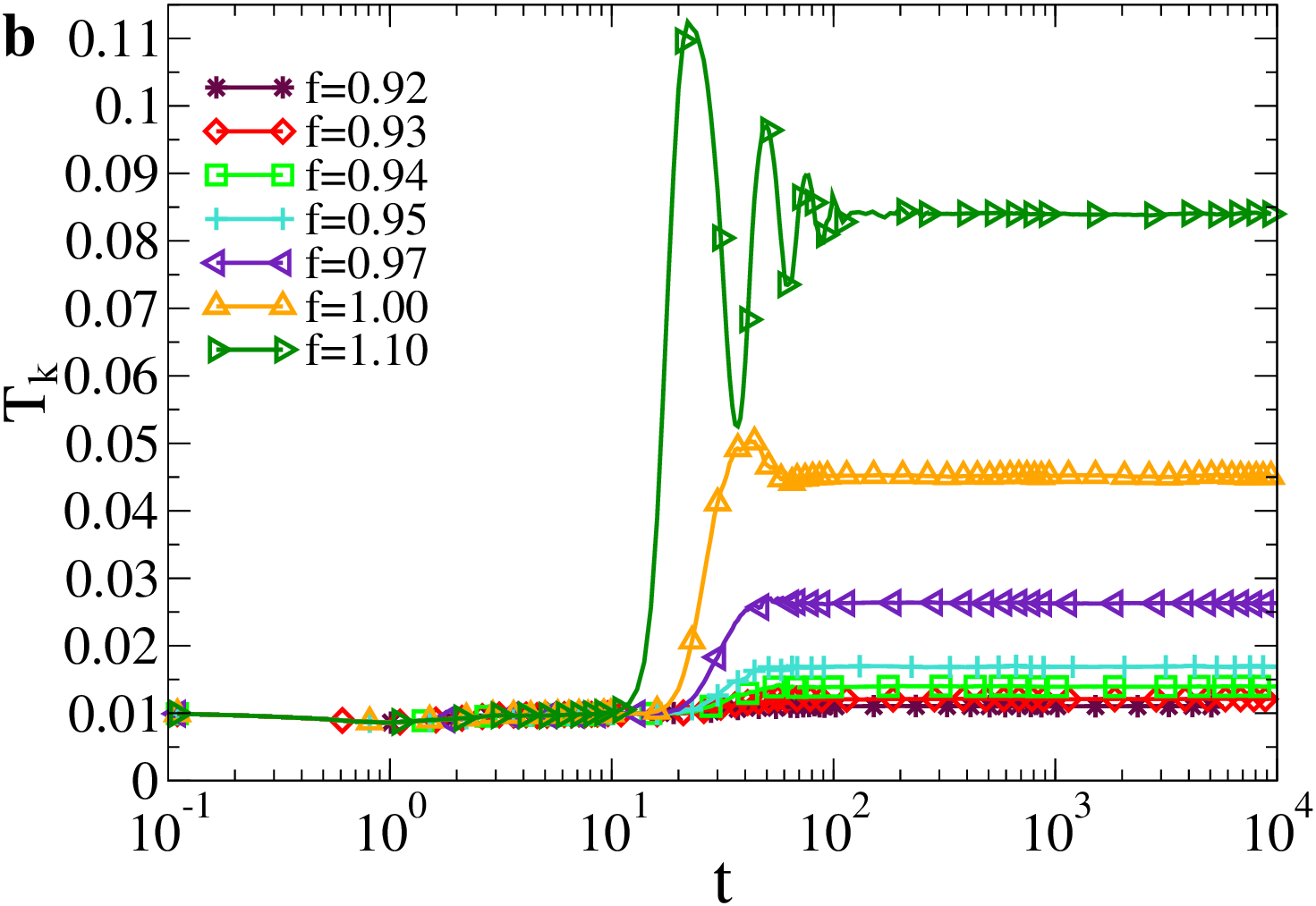} 
	
	\caption{Kinetic temperature defined as $T_k(t)=\langle \delta v^2(t)\rangle$ in the scaled
		units, \textit{vs.} time at $T=0.01$ for $\gamma_0=2$, (a) $\gamma_\alpha=3$
		($\rho=\rho_f/2$) and (b) $\gamma_\alpha=0$ (normal diffusion) at several
		values of tilting force $f$ shown in plots.  Initially, $T_k(0)=T$. The particles can
		first slightly cool down, when they start  equilibrating being localized initially in
		a potential well. Then, they arrive again at $T$ during this equilibration process and
		start  drastically heat up, when they leave the potential well, for a sufficiently
		large $f$ and $t$.  In the case of normal diffusion, $T_k(t)$ arrives at a stationary
		non-equilibrium value  $T_k^{\rm (st)}$ already for $t>100$. For $f<0.9$, the heating
		effect is almost negligible. However, the particles can become kinetically very hot
		even in the absence of memory effects. For example, at the critical tilt $f=1.00$ in
		panel (b) they are  4.5 times kinetically hotter than their surrounding. For
		strong memory effects in panel (a), the maximal $T_k^{\rm (st)}\approx 4.76\;T$
		is arrived at $f=0.94$ in this panel. Then, with a further increasing $f$, $T_k^{\rm
		(st)}$ gradually diminishes until it reaches $T$ for a very large $f$. However, in the
		memoryless case, $T_k^{\rm (st)}$ dramatically increases further with $f$ (see for
		$f=1.1$) until about $f=1.5$ (not shown) and only then gradually drops. This puzzling
		regime is left for a separate study.  $M=10^5$ particles are used for the ensemble
		averaging.}

	\label{Fig6}
\end{figure*}

\begin{figure*} 
\includegraphics[width=0.47\textwidth]{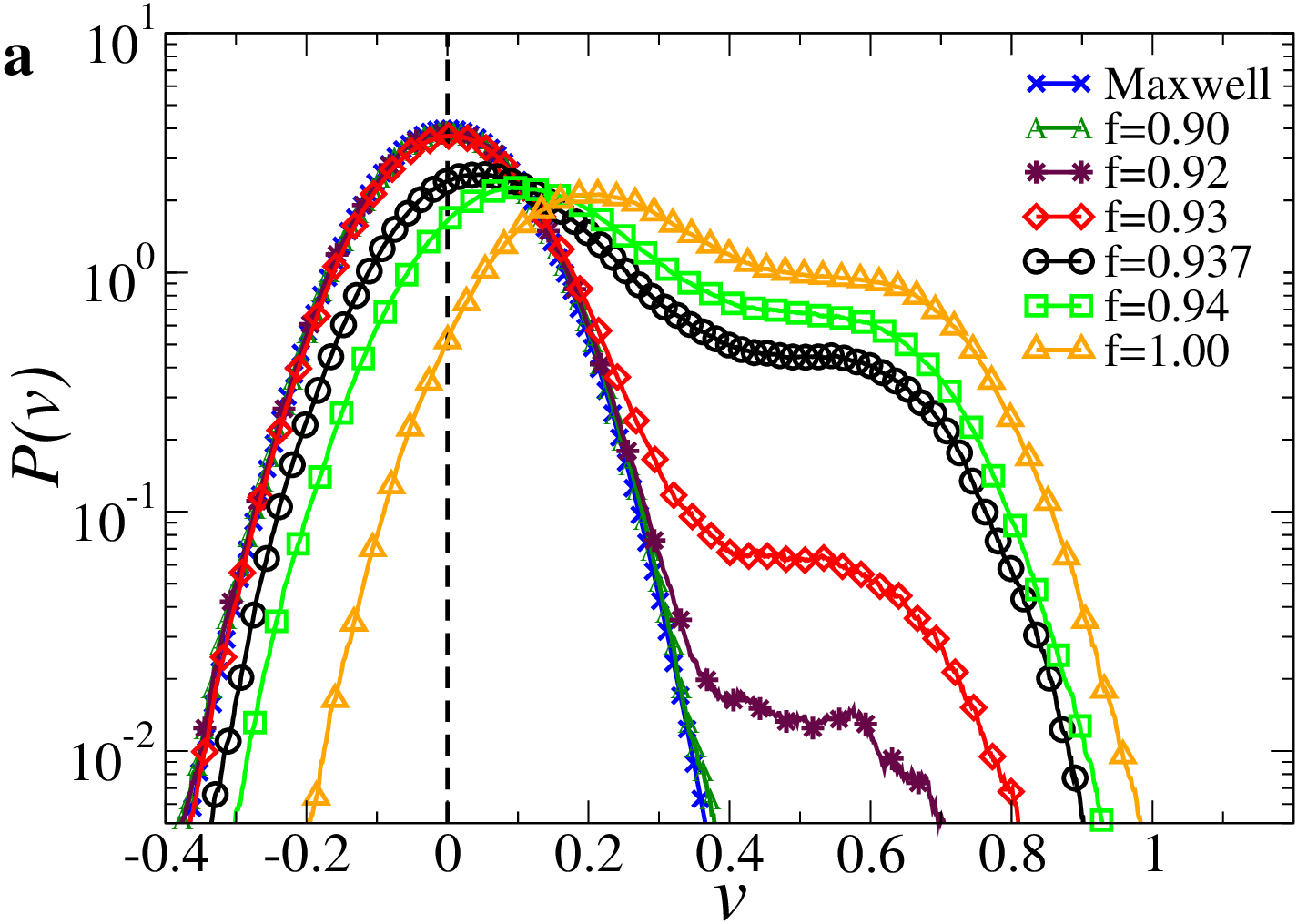}  \hspace{0.4cm}
\includegraphics[width=0.47\textwidth]{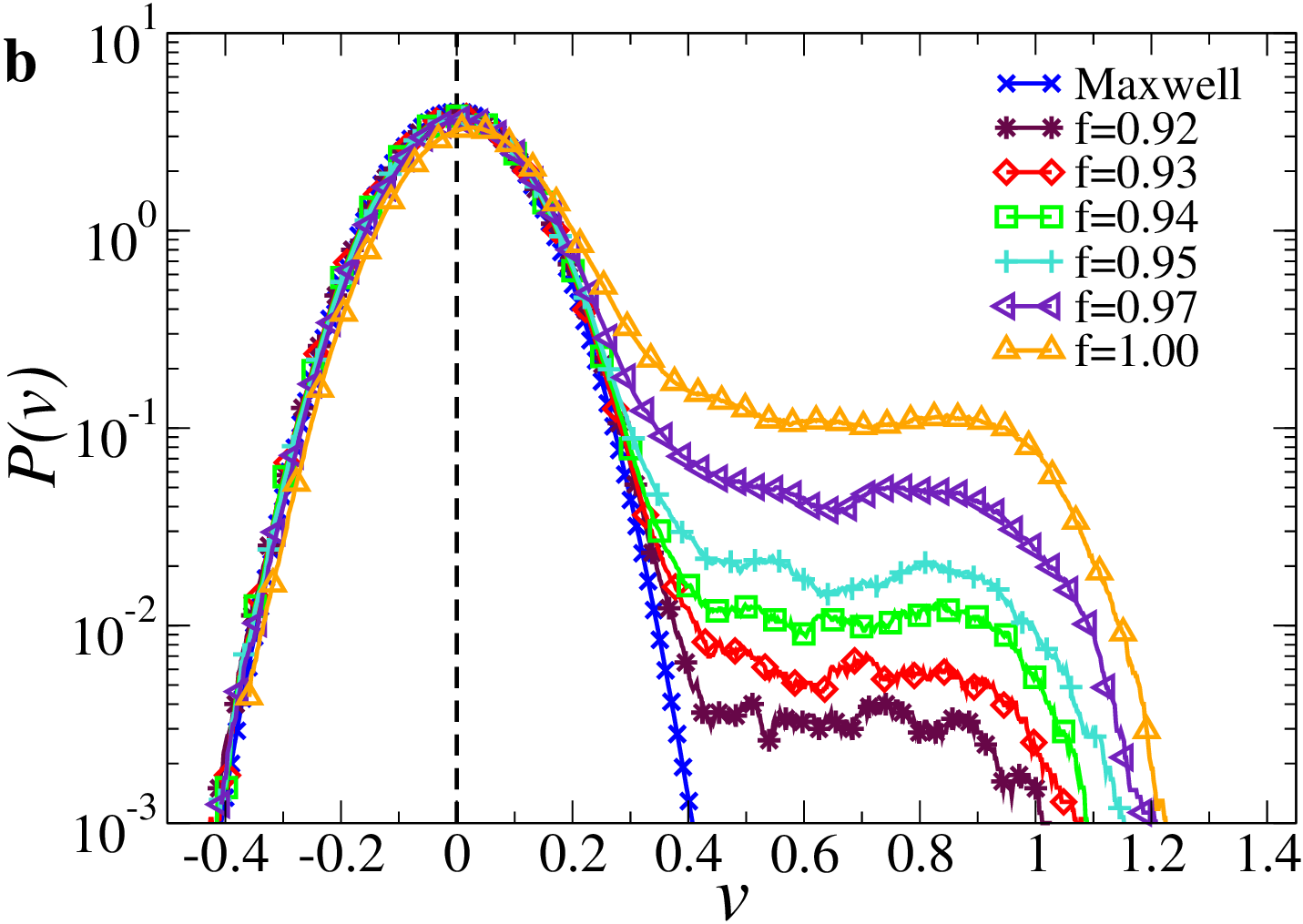} 
\caption{Velocity distribution in case of
(a) diffusion with memory and (b) memoryless diffusion at $T=0.01$, $\gamma_0=2$ and
several force values shown in plots. In panel (a), $\gamma_\alpha=3$, which corresponds to
$\rho=\rho_f/2$. Maxwell equilibrium distribution is also shown for comparison. The emerging
bimodality of $P(v)$ for a sufficiently large $f$ is obvious. For $f=0.937$ in panel (a) the
distribution is still not stationary. $M=10^5$ particles are used to produce these distributions in
each case considered. } \label{Fig7}       
\end{figure*}

In this Appendix, we discuss kinetic heating of Brownian
particles,  velocity distribution $P(v,t)$ responsible for this kinetic heating, and
their relation to enormous diffusion enhancement.

To begin with, 
in the scaling of this work, renormalized mass $m^*=1$ and $k_B=1$. Initially,
particle velocities are Maxwell-distributed, $P(v,0)=\exp[-v^2/(2T)]/\sqrt{2\pi T}$
with zero mean,  $\langle v(0)\rangle=0$, and  the variance $\langle \delta
v^2(0)\rangle =v_T^2=T$. Here and in the following, $\delta v(t)=v(t)-\langle
v(t)\rangle$. In the absence of periodic potential and for an arbitrary strong force
$f$, $P(v,t)$ after a transient time $1/\gamma_0$ is  Maxwell-distributed around the
mean value  $\langle v\rangle_{\rm st}=f/\gamma_0$.
Doing the overdamped limit at strong friction $\gamma_0\gg 1$, one assumes that
velocity distribution remains Maxwellian (shifted by $\langle v\rangle$) at the same
temperature $T$, and excludes the velocity variable from the further consideration.
This assumption becomes questionable in the case of nonlinear driven dynamics even
for a sufficiently large but finite $\gamma_0$. At the first look, dynamics in
considered tilted washboard potentials can become close to the overdamped case
already for $\gamma_0>\gamma_0^{(c)}=1.193$, as the Risken's phase diagram might
suggest, see the memoryless case in Fig. 1 of the main text.  Indeed, following this
diagram, the only critical force, which seems relevant above $\gamma_0^{(c)}$, is 
$f_c^{(1)}=1$. For $\gamma_0>\gamma_0^{(c)}$  and
$f<f_c^{(1)}$ at $T=0$ any particle will be eventually trapped in a potential well, whereas at
$f>f_c^{(1)}$ it will be running. For $\gamma_0<\gamma_0^{(c)}$  at $f<f_c^{(3)}$ and
$T=0$ all particles are eventually trapped, whereas for $f>f_c^{(3)}$ the running
solutions appear. Risken defined an effective potential for a particle using its
total energy as a variable, see Ch. 11.6 in Ref. \cite{RiskenBook}.  It displays
bistability at $f_c^{(3)}<f<f_c^{(1)}$ (in our notations) and at some critical force
$f_c^{(2)}$, both minima of that effective potential become equal. In
Refs. \cite{Marchenko12, Marchenko14}, a velocity pseudo-potential,
$V(v)=-k_BT\ln[P(v)]$, was considered instead of the Risken's potential, where $P(v)$
is the velocity distribution. It turns out to be also bistable for  sufficiently
small friction. One minimum at $v_1=0$ corresponds to the trapped particles, with
velocity being Maxwell-distributed around this minimum, and another minimum is
located at $v_2=f/\gamma_0$, with velocity also Maxwell-distributed around $v_2$.
Bi-parabolic velocity pseudo-potential with cusp at intersection of two parabolas
provides a reasonable approximation to $V(v)$ \cite{Marchenko12,Marchenko14}. This
picture of velocity bistability remains valid, upon some modifications, also in the
presence of hydrodynamic memory effects for $\gamma_0=0.1$ in
Ref. \cite{GoychukPRL19}.  

However, this simple picture breaks down even in the absence of memory effects for
sufficiently strong friction exceeding (about) $\gamma_0=0.25$; see Supplemental
Material \cite{supplPRL19} of Ref. \cite{GoychukPRL19}, which was also confirmed quite recently in
Ref. \cite{Spiechowicz20}. The numerical simulations reveal that already for
$\gamma_0=0.3$, the velocity distribution can be three-modal, see Fig. 7, (a) in the
discussed Supplemental Material \cite{supplPRL19} of Ref. \cite{GoychukPRL19}, and, especially, the
panel (c) therein, for $\gamma_0=0.7$. The running state consists, in fact, of two
velocity substates with $P(v)$ maxima at $v_{2}^{(1)}$ and $v_{2}^{(2)}$ such that
$v_{2}^{(1)}<v_2=f/\gamma_0<v_{2}^{(2)}$. Moreover, $v_2$ corresponds to the minimum
(!) and not maximum of $P(v)$, as bistable picture of
$V(v)$ \cite{Marchenko12,Marchenko14}, valid for sufficiently small $\gamma_0$
only \cite{GoychukPRL19}, can misleadingly imply. For $f\geq f_c^{(1)}$, $P(v)$
becomes bimodal in such a memoryless case because of the minimum at $v_1=0$, which
corresponds to the trapped particles, disappears -- see the panel (d) in the
discussed figure. These earlier overlooked features are important to understand the
results of this work. The velocity distribution $P(v)$ can be bimodal even for not
too large $\gamma_0$, well above the critical force $f_c^{(1)}$ (until some very
large $f$), when the tilted washboard potential does not have anymore some minima and
maxima at all. It is a great surprise overlooked until recently.

Generally, $P(v,t)$ is time-dependent for the considered nonlinear stochastic
dynamics. For a sufficiently large time, a stationary distribution $P_{\rm
st}(v)=\lim_{\to \infty} P(v,t)$ will be attained. However, this limit is not always
possible to reach in our numerics, especially in the presence of memory effects. 
Although the memory effects do not affect $P_{\rm st}(v)$  in the case of linear
dynamics, where it remains  Maxwellian, they generally essentially influence both
$P(v,t)$ and $P_{\rm st}(v)$ in the case of driven nonlinear dynamics considered.

The emerging very broad $P(v,t)$ velocity distribution, that is profoundly different from the Maxwell
distribution, means that particles become kinetically very hot. The kinetic
temperature is commonly characterized by the velocity variance $\langle \delta
v^2(t)\rangle$ 
such that $T_k(t)=\langle \delta
v^2(t)\rangle$ \cite{Brilliantov04, SieglePRL10, SiegleEPL11, Marchenko12}. For
equilibrium Maxwell distribution, $T_k(t)=T$. It is so initially in Fig.
\ref{Fig6}. As everywhere else in this paper, the particles initially are localized
at $x=0$, which is a nonequilibruim distribution within the potential well. During
the initial equilibration they are first slightly cooled, and then heated up to $T$
again. Very interesting phenomenon occurs when the particles diffuse out of the
initial potential well. Then, they can be heated up to some $T_k(t)\gg T$ reaching
finally a stationary value $T_k^{\rm (st)}$. If $f$ is far below $f_c^{(3)}$,
$T_k^{\rm (st)}=T$, as expected. For example, for $f=0.90$ in Fig. \ref{Fig7},
(a), $P_{\rm st}(v)$ is still practically Maxwellian, and no kinetic heating
occurs. However, already for $f=f_c^{(3)}=0.92$ in panel (a) of Fig.
\ref{Fig6}, $T_k^{\rm (st)}/T\approx 1.15$, i.e., $T_k$ is enhanced by about 15\%
over $T$. In the absence of memory effects, in panel (b), the enhancement is
somewhat smaller: 10\% only. The deviation from $T$ is still sufficiently small, in
both cases. The onset of $P_{\rm st}(v)$ bistability is clearly seen in Fig. 
\ref{Fig7} for $f=0.92$, also in the absence of memory effects, cf. panel
(b) therein. However, already for $f=0.93$, the kinetic temperature increases
by ca. 70\% in panel (a), Fig. \ref{Fig6}, \textit{vs.} 21\% in panel
(b). Astoundingly, a tiny further increase of force to $f=0.94$ boosts
$T_k^{\rm (st)}$ to about $T_k^{\rm (st)}=4.76\;T$ in panel (a), which is the
maximal stationary value therein. This sharp increase should be contrasted with a
still small increase in the memoryless case in Fig. \ref{Fig6}, (b). Hence,
the discussed sharp increase is caused by the memory effects indeed.  With a further
increase of $f$, $T_k^{\rm (st)}$ gradually diminishes and for a very large force it
finally drops down to $T$ again (not shown). The force range, where the Brownian
particles become hot under constant driving is surprisingly large. It must be
mentioned also that for $f=f_c^{(2)}=0.937$ and $f=0.938$, $T_k(t)$ still did not
reach the stationary value $T_k^{\rm (st)}$ in panel (a) of Fig. \ref{Fig6}.
The corresponding distribution $P(v)$ for $f=0.937$ in Fig. \ref{Fig7}, (a)
is still not stationary. This still increasing $T_k(t)$ in Fig. \ref{Fig6}
correlates with transient superdiffusion which lasts until the end of simulations in
these cases, see Fig. 3, (a) of the main text. Maximum of the corresponding
asymptotic diffusion coefficient estimated in Fig. 4, (a) of the main text
indeed seems to be associated with the maximum of $T_k^{\rm (st)}$ \textit{vs.} $f$,
as in the case of small $\gamma_0$ \cite{GoychukPRL19}. However, here some profound
warnings are due. First, even if $T_k(t)$ is already saturated for $f=0.94$ in the
discussed case, cf. panel (a) of Fig. \ref{Fig6}, the corresponding
diffusional behavior in Fig. 3, (a) of the main text is still transient, and,
unexpectedly, displays subdiffusion. For a much smaller $\gamma_0=0.1$ in
Ref. \cite{GoychukPRL19} the emergence of such a transient subdiffusion regime was
connected with the regime like one for $f=0.95$ in panel (a) of Fig.
\ref{Fig6}, when $T_k(t)$ drops gradually to  $T_k^{\rm (st)}$ after reaching a
maximum. Indeed, also for this case, a transient subdiffusive behavior is still
detectable (not shown). However, already for $f=1$, one cannot find such a regime,
even if the corresponding non-monotonous behavior of $T_k(t)$ in panel (a) of
Fig. \ref{Fig6} might imply it. It becomes simply too short to be detectable.
Second, in the memoryless case, the maximum of $T_k^{\rm (st)}$ does not correspond
to the maximum of $D$.  Indeed, in Fig. \ref{Fig6}, (b), $T_k^{\rm (st)}$ is
the largest for $f=1.1$ and not for $f=1.0$ which corresponds to the maximum of $D$
in Fig. 4, (a) of the main text. In other words, increase of $f$ beyond
$f_c^{(1)}$ in the memoryless case leads to a further increase of $T_k^{\rm (st)}$,
cf. Fig. \ref{Fig6}, (b), whereas $D$ already starts to diminish, cf. Fig.
4, (a) of the main text. Here, a fascinating novel phenomenon emerges, which
seems to be completely overlooked thus far and which we defer to a separate study.
Namely, the \textit{stationary } velocity autocorrelation function (VACF) starts, to
our great surprise, rapidly oscillate in time (not shown). These oscillations
correlate with \textit{transient} $T_k(t)$ oscillations like ones for $f=1.1$ in Fig.
\ref{Fig6}, (b), which are also quite surprising.   Since the diffusion
coefficient is integral of VACF\cite{KuboBook,ChaikinBook}, $D$ declines with $f$
despite $T_k^{\rm (st)}$ still grows. This doubly unusual phenomenon indicating that
the inertial effects can remain significant for appreciably large $\gamma_0$, even in
the absence of hydrodynamic memory effects, will be studied in a separate work. It is
especially striking and surprising because the diffusion enhancement in this work is
already well described by the results of overdamped theory in the corresponding case.

The behavior of $P(v)$ in Fig. \ref{Fig7} also deserves a separate discussion.
Notice that at $f=0.90$ in panel (a), $P(v)$ is still pretty well described by
the equilibrium Maxwell distribution, with the center which is still practically not
shifted. For $f=0.92$, the emerging velocity bistability becomes perspicuous.
However, the first maximum is still centered at $v=0$ corresponding to the trapped
states. Nevertheless, already for $f=0.93$, this maximum shifts slightly to the
running states, indicating that for a finite temperature, the trapped states are
destabilized by thermal fluctuations for $f>f_c^{(3)}$, in the case considered, which
is a remarkable feature. Likewise, already at $f=1$ in the case of memoryless
diffusion, the first maximum of $P(v)$ shifts to the running states, cf. Fig.
\ref{Fig7}, (b). One might expect it because, at $f=1$, the potential minima
vanish overall. However, such a distinct shift at $f=0.937$ in panel (a) is
quite surprising.

\bibliographystyle{PRX}
\bibliography{prx}

\end{document}